\documentclass[prd,11pt,nofootinbib,superscriptaddress,preprintnumbers,floatfix]{revtex4-1}
\usepackage{amsmath,amssymb}
\usepackage{graphicx}
\usepackage{subfigure}
\usepackage{hyperref}
\usepackage{nicefrac}
\usepackage{color}
\usepackage{dsfont}

\newcommand{\pt}{\mathrm{pt}}

\newcommand{\gev}{~\mathrm{GeV}}

\newcommand{\fig}[1]{Figure~\ref{#1}}

\newcommand{\eq}[1]{Eq.~(\ref{#1})}

\begin{document}

\title{The pion vector form factor from Lattice QCD at the physical point}

\newcommand\cypa{Computation-based Science and Technology Research Center, The Cyprus Institute,\\ PO Box 27456, 1645 Nicosia, Cyprus}
\newcommand\cypb{Department of Physics, University of Cyprus, PO Box 20537, 1678 Nicosia, Cyprus}
\newcommand\cypc{The Cyprus Institute, 20 Kavafi Str., Nicosia 2121, Cyprus}
\newcommand\ber{Albert Einstein Center for Fundamental Physics, University of Bern, 3012 Bern, Switzerland}
\newcommand\bn{HISKP and BCTP, Rheinische Friedrich-Wilhelms Universit\"at Bonn, 53115 Bonn, Germany}
\newcommand\fer{Centro Fermi - Museo Storico della Fisica e Centro Studi e Ricerche "Enrico Fermi",\\ Compendio del Viminale, Piazza del Viminale 1, 00184 Roma, Italy}
\newcommand\gre{Theory Group, Lab. de Physique Subatomique et de Cosmologie, 38026 Grenoble, France}
\newcommand\infn{Istituto Nazionale di Fisica Nucleare, Sezione di Roma Tre,\\ Via della Vasca Navale 84, I-00146 Rome, Italy}
\newcommand\NIC{John von Neumann Institute for Computing (NIC), DESY, Platanenallee 6, 15738 Zeuthen, Germany}
\newcommand\rmii{Dipartimento di Fisica, Universit{\`a} di Roma Tor Vergata e Sezione INFN di Roma Tor Vergata,\\ Via della Ricerca Scientifica 1, I-00133 Roma, Italy}
\newcommand\wup{Fakult\"at f\"ur Mathematik und Naturwissenschaften, Bergische Universit\"at Wuppertal,\\ 42119 Wuppertal, Germany}

\author{ C.~Alexandrou}\affiliation{\cypb}\affiliation{\cypc}
\author{ S.~Bacchio}\affiliation{\cypb}\affiliation{\wup}
\author{P.~Dimopoulos}\affiliation{\rmii}\affiliation{\fer}
\author{J.~Finkenrath}\affiliation{\cypa}
\author{R.~Frezzotti}\affiliation{\rmii}
\author{K.~Jansen}\affiliation{\NIC}
\author{B.~Kostrzewa}\affiliation{\bn}
\author{M.~Mangin-Brinet}\affiliation{\gre}
\author{F.~Sanfilippo}\affiliation{\infn}
\author{S.~Simula}\affiliation{\infn}
\author{C.~Urbach}\affiliation{\bn}
\author{U.~Wenger}\affiliation{\ber}
%\collaboration{ETM Collaboration}

\begin{abstract}
\begin{center}
\includegraphics[draft=false,width=.2\linewidth]{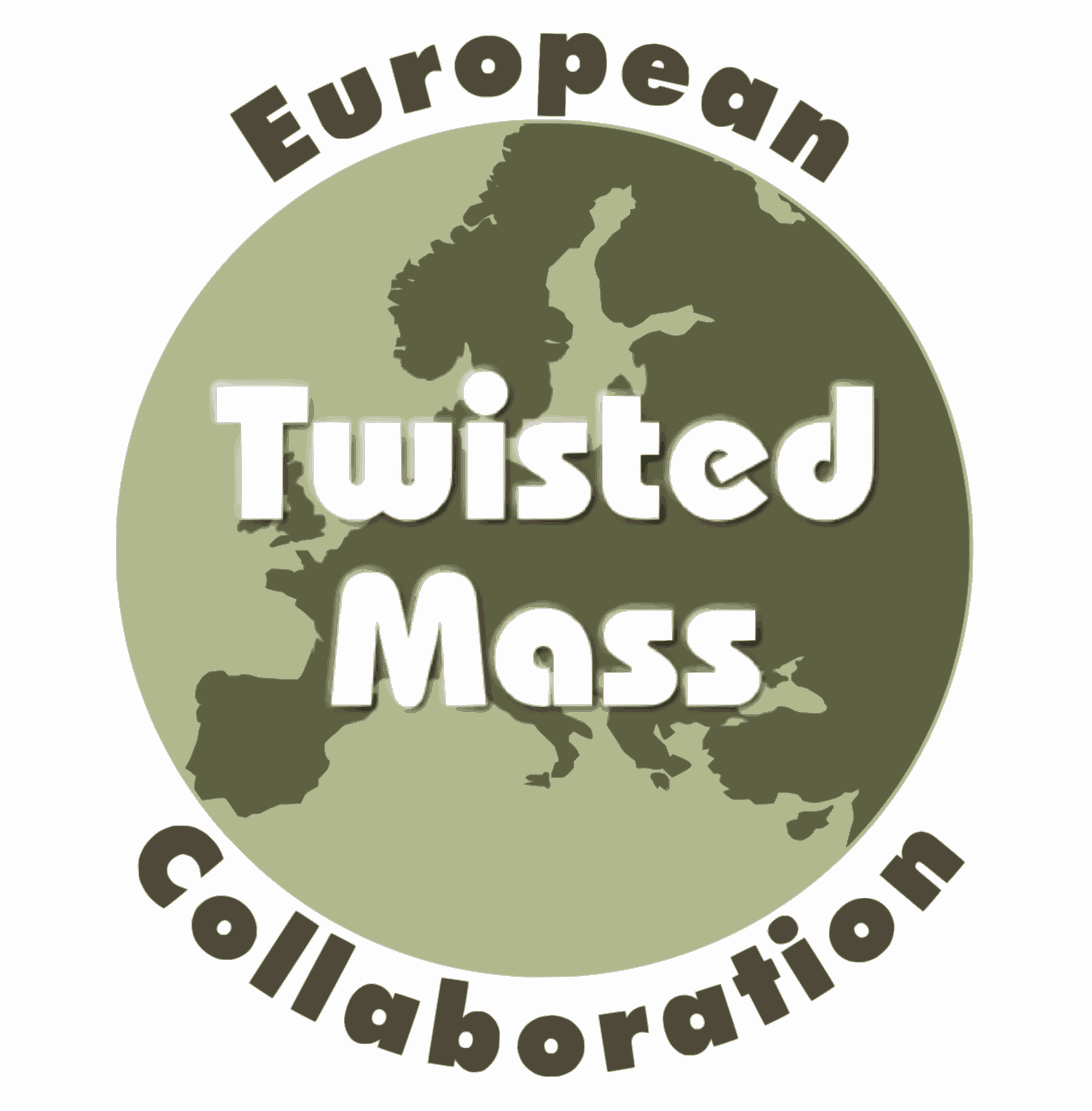}
\end{center}
We present an investigation of the electromagnetic pion form factor, $F_\pi(Q^2)$, at small values of the four-momentum transfer $Q^2$ ($\lesssim 0.25\gev^2$), based on the gauge configurations generated by European Twisted Mass Collaboration with $N_f = 2$ twisted-mass quarks at maximal twist including a clover term.
Momentum is injected using non-periodic boundary conditions and the calculations are carried out at a fixed lattice spacing ($a \simeq 0.09$ fm) and with pion masses equal to its physical value, 240 MeV and 340 MeV.
Our data are successfully analyzed using Chiral Perturbation Theory at next-to-leading order in the light-quark mass.
For each pion mass two different lattice volumes are used to take care of finite size effects.
Our final result for the squared charge radius is $\langle r^2 \rangle_\pi = 0.443~(29)$ fm$^2$, where the error includes several sources of systematic errors except the uncertainty related to discretization effects. 
The corresponding value of the SU(2) chiral low-energy constant $\overline{\ell}_6$ is equal to $\overline{\ell}_6 = 16.2 ~ (1.0)$.
\end{abstract}

\maketitle

\clearpage

%%%%%%%%%%%%%%%%%%%%%%%%%%%%%%%%%%%%%%%%%%%%%%%%
\section{Introduction}
\label{sec:intro}

The investigation of the physical properties of the pion, which is the lightest 
bound state in Quantum Chromodynamics (QCD), can provide crucial information on
the way low-energy dynamics is governed by the quark and gluon degrees of 
freedom.
In this respect for space-like values of the squared four-momentum transfer, 
$Q^2 \equiv - q^2 \geq 0$, the electromagnetic (e.m.) form factor of the pion, $F_\pi(Q^2)$,
parametrizes how the pion deviates from a point particle when probed
electromagnetically, thus giving insight on the distribution of its
charged constituents.
At momentum transfer below the scale of chiral symmetry breaking ($Q^2 \lesssim 
1\gev^2$) the pion form factor represents therefore an important test of 
non-perturbative QCD.

It is well known that for $Q^2 \lesssim 0.5 \div 1\gev^2$ the experimental data on the pion form factor \cite{Blok:2008jy,Huber:2008id,Amendolia:1986wj} can be 
reproduced qualitatively by a simple monopole ansatz inspired by the Vector 
Meson Dominance (VMD) model with the contribution from the lightest vector 
meson ($M_\rho \simeq 0.77\gev$) only. 
This is not too surprising in view of the fact that in the time-like region 
the pion form factor is dominated by the $\rho$-meson resonance.

An interesting issue is the quark mass dependence of the pion form factor that can 
be addressed by SU(2) Chiral Perturbation Theory (ChPT) known at both next-to-leading 
(NLO) \cite{Gasser:1983yg} and next-to-next-to-leading (NNLO) \cite{Bijnens:1998fm} orders. 
The determination of $F_\pi(Q^2)$ from lattice QCD simulations provides therefore 
an excellent opportunity for the study of chiral logarithms.
The latter are particularly important in the case of the squared pion charge radius $r_\pi^2$, 
i.e.~the slope of pion form factor at $Q^2 =0$.
This means also that a controlled extrapolation to the physical point
is a delicate endeavour, such that one would ideally like to perform
the computation directly at the physical pion mass.

Initial studies of the pion form factor using lattice QCD in the quenched approximation date back to the late 80's \cite{Martinelli:1987bh,Draper:1988bp} giving strong support to the vector-meson dominance hypothesis at low $Q^2$. 
Studies of $F_\pi(Q^2)$ employing unquenched simulations have been carried out in Refs.~\cite{Brommel:2006ww,Frezzotti:2008dr,Boyle:2008yd,Aoki:2009qn,Nguyen:2011ek,Brandt:2013dua,Fukaya:2014jka,Aoki:2015pba} using pion masses above the physical one and adopting ChPT as a guide to extrapolate the lattice results down to the physical pion point.
Recently a computation of $F_\pi(Q^2)$ at the physical pion mass has been provided in Ref.~\cite{Koponen:2015tkr}.

In this work we present a determination of the pion form factor using the gauge configurations generated in Ref.~\cite{Abdel-Rehim:2015pwa} by the European Twisted Mass Collaboration (ETMC) with $N_f = 2$ twisted-mass quarks at maximal twist, which guarantees the automatic ${\cal{O}}(a)$-improvement~\cite{Frezzotti:2003ni}.
The calculations are carried out at a fixed lattice spacing ($a \simeq 0.09$ fm) and with pion masses equal to its physical value, 240 MeV and 340 MeV.
Momentum is injected using non-periodic boundary conditions in order to get values of $Q^2$ between $\simeq 0.01 \gev^2$ and $\simeq 0.25 \gev^2$.
It will be shown that our data can be successfully analyzed using SU(2) ChPT at NLO without the need of the scale setting.

Our final result for the squared pion charge radius is 
 \begin{equation}
      \langle r^2 \rangle_\pi = 0.443~(29)~\mbox{fm}^2 ~ ,
      \label{eq:rpi2_final}
 \end{equation}
where the error includes several sources of systematic errors except the uncertainty related to discretization effects. 
The corresponding value of the NLO SU(2) low-energy constant (LEC) $\overline{\ell}_6$ is equal to
 \begin{equation}
     \overline{\ell}_6 = 16.2 ~ (1.0) ~ .
 \end{equation}
Our result (\ref{eq:rpi2_final}) is obtained at a fixed value of the lattice spacing and therefore the continuum limit still needs to be evaluated. 
We note that discretization effects in our calculations of the pion form factor start at order ${\cal{O}}(a^2)$ (see Section \ref{sec:Fpi}) and that our finding (\ref{eq:rpi2_final}) is consistent with the experimental value $\langle r^2 \rangle_\pi^{exp.} = 0.452~(11)$ fm$^2$ from PDG~\cite{Olive:2016xmw}.
This suggests that the impact of discretization effects on our result (\ref{eq:rpi2_final}) could be small with respect to the other sources of uncertainties.

The plan of the paper is as follows.
In Section \ref{sec:actions} we describe the lattice setup adopted in this work, while the procedures adopted to extract the pion form factor from appropriate ratios of 3- and 2-point correlators are discussed in Section \ref{sec:Fpi}.
The lattice data for the pion form factor $F_\pi(Q^2)$ are presented in Section \ref{sec:results} and in the Appendix, while our fitting procedures based on ChPT are described in Section \ref{sec:method}.
The results of the extrapolations to the physical point and to the infinite lattice volume are collected in Section \ref{sec:extrapolations}.
Our conclusions are summarized in Section \ref{sec:conclusions}.

%%%%%%%%%%%%%%%%%%%%%%%%%%%%%%%%%%%%%%%%%%%%%%%%
\section{Lattice action}
\label{sec:actions}

The results presented in this paper are based on the gauge
configurations generated in Ref.~\cite{Abdel-Rehim:2015pwa} by the ETMC with Wilson clover twisted mass
quark action at maximal   
twist~\cite{Frezzotti:2000nk}, employing the Iwasaki gauge
action~\cite{Iwasaki:1985we}. 
The measurements are performed on $N_f = 2$ ensembles with
pion mass at its physical value, 240~MeV and 340~MeV,
respectively. The lattice spacing is $a \simeq 0.0914~(15)~\mathrm{fm}$ for
all the ensembles~\cite{Abdel-Rehim:2015pwa}.  
In Table~\ref{tab:setup} we list the 
ensembles with the relevant input parameters, the lattice volume and
the number of configurations used. More details about the ensembles are
presented in Ref.~\cite{Abdel-Rehim:2015pwa}. 

\begin{table}[htb!]
 \centering
 \begin{tabular*}{.9\textwidth}{@{\extracolsep{\fill}}lcccccc}
  \hline\hline
  ensemble & $\beta$ & $c_{\mathrm{sw}}$ &$a\mu_\ell$  &$(L/a)^3\times T/a$ & $N_\mathrm{conf}$ & $aM_\pi$ \\ 
  \hline\hline
  $cA2.09.64$ &2.10 &1.57551 &0.009  &$64^3\times128$ & $360$ & $0.06204 ~~ (6)$ \\
  $cA2.09.48$ &2.10 &1.57551 &0.009  &$48^3\times96$ & $615$ & $0.06216 ~~ (8)$ \\
  $cA2.30.48$ &2.10 &1.57551 &0.030  &$48^3\times96$ & $345$ & $0.11198 ~~ (9)$ \\
  $cA2.30.24$ &2.10 &1.57551 &0.030  &$24^3\times48$ & $300$ & $0.11567 ~ (85)$ \\
  $cA2.60.32$ &2.10 &1.57551 &0.060  &$32^3\times64$ & $330$ & $0.15773 ~ (25)$ \\
  $cA2.60.24$ &2.10 &1.57551 &0.060  &$24^3\times48$ & $270$ & $0.15861 ~ (83)$ \\
  \hline\hline
 \end{tabular*}
 \caption{\it \footnotesize The gauge ensembles used in this study. The labelling of
   the ensembles follows the notations in
   Ref.~\cite{Abdel-Rehim:2015pwa}. In addition to the relevant input
   parameters we give the lattice volume $(L/a)^3\times T/a$, the
   number of evaluated configurations $N_\mathrm{conf}$ and 
   the pion mass $M_\pi$ in lattice units (with statistical error).}
 \label{tab:setup}
\end{table}

Both the sea and valence quarks are described by the Wilson clover twisted mass
action. The Dirac operator for the light quark doublet consists of the
Wilson twisted mass Dirac operator~\cite{Frezzotti:2000nk} combined
with the clover term, namely in the so-called physical basis
\begin{equation}
  \label{eq:Dlight}
  D_\ell = D -i\gamma_5\tau_3 \left[W_\mathrm{cr} + 
  \frac{i}{4}c_\mathrm{sw}\sigma^{\mu\nu}\mathcal{F}^{\mu\nu}\right] + \mu_\ell
\end{equation}
where $D = \gamma_\mu(\nabla^\ast_\mu+\nabla_\mu)/2$, $\nabla_\mu$ and
$\nabla_\mu^\ast$ are the forward and backward lattice covariant
derivatives, and $W_\mathrm{cr} = -(a/2)\nabla_\mu^\ast\nabla_\mu +
m_\mathrm{cr}$ with $m_\mathrm{cr}$ being the critical mass. Moreover, $\mu_\ell$ is
the average up/down (twisted) quark mass, $a$ is the lattice spacing
and $r=1$ the Wilson parameter. The operator $D_\ell$ acts on a flavour doublet
spinor $\psi = (u,d)^T$. Finally, $c_\mathrm{sw}$ is the so-called
Sheikoleslami-Wohlert improvement coefficient
\cite{Sheikholeslami:1985ij} multiplying the clover term. 
In our case the latter is not used for $\mathcal{O}(a)$ improvement but serves to
significantly reduce the effects of isospin
breaking~\cite{Abdel-Rehim:2015pwa}. 

The critical mass has been determined as described in
Refs.~\cite{Chiarappa:2006ae,Baron:2010bv}. This guarantees 
that all physical observables can be extracted from
lattice estimators that are O(a) improved by symmetry 
\cite{Frezzotti:2003ni}, which is one of the main advantages of the
Wilson twisted mass formulation of lattice QCD.

%%%%%%%%%%%%%%%%%%%%%%%%%%%%%%%%%%%%%%%%%%%%%%%%
\section{The pion form factor}
\label{sec:Fpi}

The pion form factor can be computed from the matrix elements of the e.m.~vector current 
\begin{equation}
  V_\mu(x) = \frac{2}{3} \bar{u}(x) \gamma_\mu u(x) - \frac{1}{3} \bar{d}(x) \gamma_\mu d(x) 
  \label{eq:Vmu}
\end{equation}
between pion states, yielding
\begin{equation}
  \left\langle \pi^+(\vec{p^\prime}) | V_\mu(0) | \pi^+( \vec{p}) \right\rangle
  = ( p_\mu^\prime + p_\mu ) \, F_\pi(Q^2) \, ,
  \label{eq:Fpi}
\end{equation}
where $q_\mu = ( p_\mu - p_\mu^\prime )$  is the 4-momentum transfer and $Q^2 \equiv -q^2$.
As detailed in Ref.~\cite{Frezzotti:2008dr}, up to discretization effects of order ${\cal{O}}(a^2)$ it is enough to compute in \eq{eq:Fpi} only the connected insertion of the single flavor current $\bar{u}(x) \gamma_\mu u(x)$ with unitary charge.

\begin{figure}[htb!]
  \centering
  \includegraphics[width=.75\linewidth]{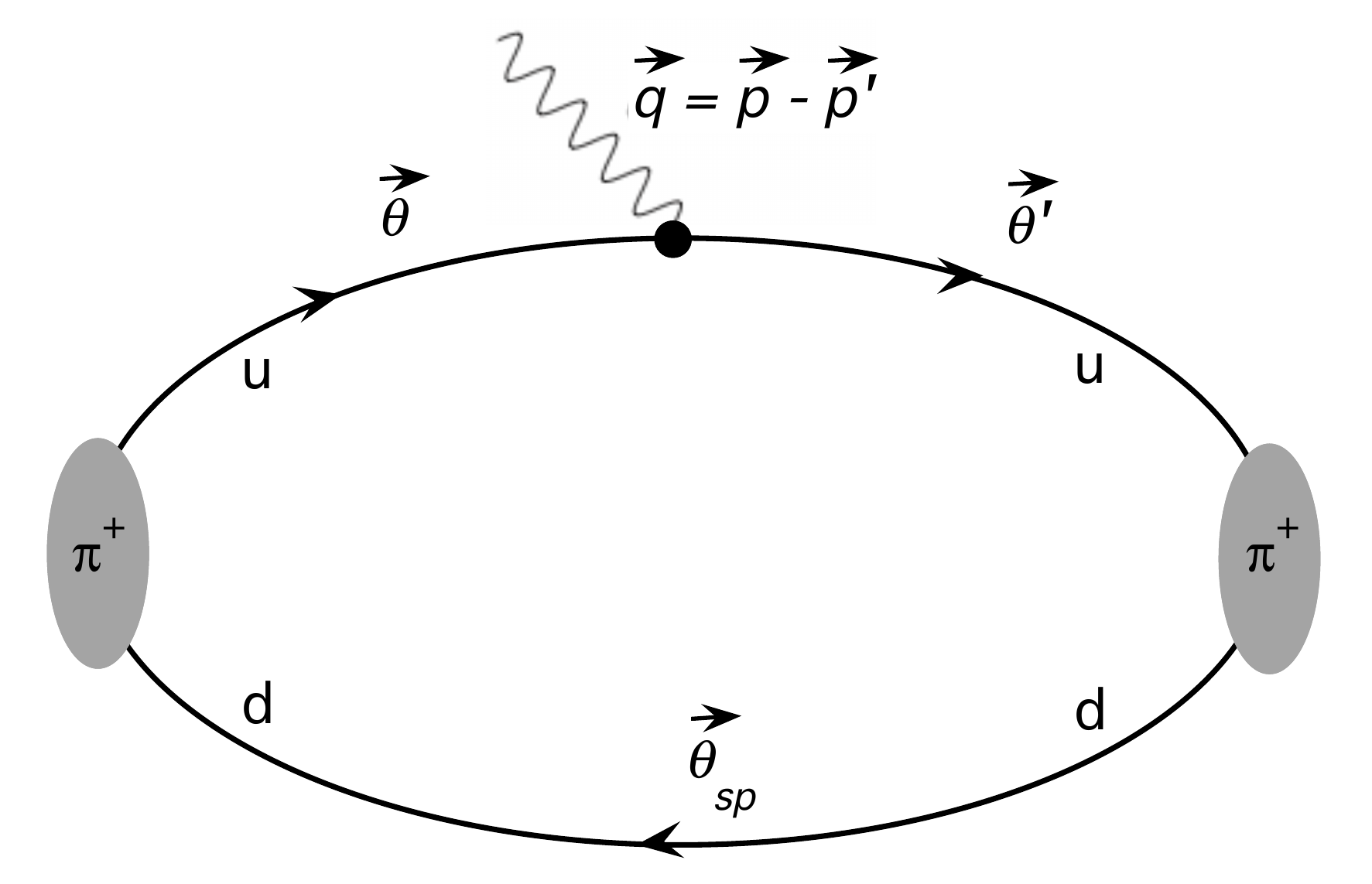}
  \vspace{-0.5cm}
  \caption{\it \footnotesize Non-periodic boundary conditions in the pion three-point
    function for arbitrary quark momenta.}
  \label{fig:pionthreep}  
\end{figure}

Working in Euclidean space-time, we can access the region of space-like
momentum transfer, $Q^2 = - q^2 >0$, by evaluating ratios of pion two-point
and three-point functions with the vector current insertion.
To inject arbitrary momenta, we make use of \emph{non-periodic} boundary
conditions (BCs)
\cite{Bedaque:2004kc,deDivitiis:2004kq,Guadagnoli:2005be} on the quark
fields. 
Enforcing $\psi(x+\vec{e}_i L) = e^{2\pi i\theta_i} \psi(x)$ on the quark
field $\psi$, changes the momentum quantisation 
condition in finite volume to $p_i = \frac{2\pi \theta_i}{L} + \frac{2\pi
  n_i}{L}$.
This is depicted in \fig{fig:pionthreep} for the pion three-point
function with independent values of the vector $\vec{\theta}$ for the three
quark lines.
Since the ETMC gauge ensembles have been produced by imposing antiperiodic BCs in time, the same conditions are applied also to the valence quarks choosing $2\pi \theta_0 / T = \pi / T$.
Moreover, the use of different BCs in space for sea and valence quarks produces unitarity violating finite volume effects, which are however exponentially small~\cite{Sachrajda:2004mi,Bedaque:2004ax,Flynn:2005in}.

For the case of twisted mass quarks, this setup was first studied in
Ref.~\cite{Frezzotti:2008dr} in the Breit frame ($\vec{p^\prime} = -\vec{p}$),
which results in a squared 4-momentum transfer independent 
of the pion mass, viz.
\begin{equation*}
  Q^2 \equiv - q^2 =  \left| \vec{p} - \vec{p^\prime} \right|^2 - 
                                  \left[ E_\pi(\vec{p}) - E_\pi(\vec{p^\prime}) \right]^2 
                            = 4\left| \vec{p} \right|^2 \, .
\end{equation*}
To obtain Breit frame kinematics with non-periodic BCs, we set $\vec{\theta}^\prime =
- \vec{\theta}$ and $\vec{\theta}_{sp}=0$ (see Fig.~\ref{fig:pionthreep}). 
In this work the spatial components of the vector $\vec{\theta}$ are always chosen to be equal each other, i.e.~$\vec{\theta} = \{ \theta, \theta, \theta \}$.

Following Ref.~\cite{Frezzotti:2008dr} the required correlation
functions can be evaluated efficiently through the usage of the
so-called one-end-trick combined with spatial all-to-all propagators
from stochastic time-slice sources and the sequential propagator
method for the insertion (see Ref.~\cite{Baron:2007ti} for the idea
first applied to moments of pion parton distribution functions). 
Since the spatial matrix elements of the vector current are vanishing in
the Breit frame, we have to compute the following correlation
functions
\begin{align}
  \label{eq:cor2}
  C^{2\pt}(t,\vec{p}) & = \sum_{x,z} \left\langle O_\pi(x)
  O^\dagger_\pi(z) \right\rangle \delta_{t,t_x-t_z} e^{-i\vec{p}\cdot
    (\vec{x}-\vec{z})} \\
  \label{eq:cor3}
  C^{3\pt}_0(t,t',\vec{p},-\vec{p}) & = \sum_{x,y,z} \left\langle
  O_\pi(y) V_0(x) O^\dagger_\pi(z) \right\rangle
                \delta_{t,t_x-t_z} \delta_{t',t_y-t_z} e^{-i
                  \vec{p}\cdot(\vec{x}-\vec{z}) -
                  i\vec{p}\cdot(\vec{x}-\vec{y}) } \, ,
\end{align}
where $V_0(x) = \bar{u}(x) \gamma_0 u(x)$ is the temporal component of the local vector current, $O_\pi(x)=\bar{d}(x)\gamma_5 u(x)$ is the interpolating operator annihilating the $\pi^+$, $t$ is the time distance between the vector current insertion and the source and $t'$ is the time distance between the sink and the source.

As it has been shown in Ref.~\cite{Frezzotti:2003ni}, the calculation of correlation functions of globally parity invariant operators is automatically ${\cal{O}}(a)$ improved at maximal twist.
Thus, for non-vanishing values of the spatial momenta the ${\cal{O}}(a)$ terms can be eliminated by appropriate averaging of the correlation functions over initial and final momenta of opposite sign.
Using the invariance of our lattice formulation under an even number of space or time inversions and under charge conjugation as well as the $\gamma_5$-hermiticity property, one gets that: ~ i) the correlators (\ref{eq:cor2}) and (\ref{eq:cor3}) are real, and ~ ii) $C^{2\pt}(t, \vec{p}) = C^{2\pt}(t, -\vec{p})$ and $C_0^{3\pt}(t, t^\prime, \vec{p}, -\vec{p}) = C_0^{3\pt}(t, t^\prime, -\vec{p}, \vec{p})$.
Thus, we have $E_\pi(\vec{p}) = E_\pi(-\vec{p})$ and the discretization effects in both $C^{2\pt}(t, \vec{p})$ and $C_0^{3\pt}(t, t^\prime, \vec{p}, -\vec{p})$ start automatically at order ${\cal{O}}(a^2)$.

Taking the appropriate limits with $T$ being the time extent of the
lattice, one obtains in the Breit frame
\begin{align}
  \lim_{\substack{t\to\infty \\ T \to \infty}} C^{2\pt}(t,\vec{p}) &
  \to \frac{G_\pi^2}{2E_\pi(\vec{p})} e^{-E_\pi(\vec{p})t} \\
  \lim_{\substack{t\to\infty \\ (t'-t)\to\infty \\ T \to \infty}}
  C^{3\pt}_0(t,t',\vec{p},-\vec{p}) & \to
  \frac{G_\pi^2}{2E_\pi(\vec{p})2E_\pi(\vec{p})}
        \left\langle \pi^-(\vec{p}) | V_0 | \pi^+(\vec{p})
        \right\rangle e^{-E_\pi(\vec{p})t} e^{-E_\pi(\vec{p})(t'-t)}
        \, ,
\end{align}
where $G_\pi^2$ is the amplitude of the 2-point correlation function. Since we
work from now on exclusively in the Breit frame, we will drop the
second momentum argument and write
\[
C^{3\pt}_0(t,t',\vec{p},-\vec{p})\ \equiv\ C^{3\pt}_0(t,t',\vec{p})\,.
\]
Now, we can construct the ratio
\begin{equation}
  \label{eq:ratio}
  R(t, t', \vec p)\ =\ \frac{C^{3\pt}_0(t,t',\vec{p})}{C^{2\pt}(t',\vec{p})}\,,
\end{equation}
which has the following combined limit
\begin{equation*}
  \lim_{\substack{t\to\infty \\ (t'-t)\to\infty \\ T\to\infty}}
  R(t, t', \vec p) \to 
        \frac{\left\langle \pi^+(-\vec{p}) | V_0 | \pi^+(\vec{p})
          \right\rangle}{2E_\pi(\vec{p})} = \frac{1}{Z_V} F_\pi(Q^2)
        \,.
\end{equation*}
To extract $F_\pi(Q^2)$, we compute the renormalisation constant of
the vector current, $Z_V$, from the ratio of the two and three-point
functions at zero momentum transfer and the known normalisation
$F_\pi(0)=1$, which implies
\begin{equation}
  \lim_{\substack{t \to \infty \\ (t' - t) \to \infty \\ T \to
      \infty}} \,
  \frac{C^{2\pt}(t,\vec{0})}{C^{3\pt}_0(t,t',\vec{0})} \to Z_V \, .
  \label{eq:ZV}
\end{equation}

In practice, Eqs.~(\ref{eq:cor2}-\ref{eq:cor3}) are evaluated by first generating stochastic
sources $\xi_r^{a, \alpha}(\vec x, t)$ ($r = 1, ..., N$) at a single (randomly chosen) time-slice, that for ease of notation we conventionally put in what follows at $t_{\rm source} = 0$, namely 
\begin{equation}
  \begin{split}
    \lim_{N\to\infty} &\frac{1}{N}\sum_{r=1}^N \xi_r^{a, \alpha}(\vec
    x , 0)^\star\cdot \xi_r^{b,
      \beta}(\vec y , 0) =
    \delta_{a,b}\delta_{\alpha,\beta}\delta_{\vec x, \vec y}\,,\\
    \lim_{N\to\infty} &\frac{1}{N}\sum_{r=1}^N \xi_r^{a, \alpha}(\vec
    x, t) \cdot \xi_r^{b, \beta}(\vec y, t) = 0\,.
  \end{split}
\end{equation}
Here, $a$($b$) and $\alpha$($\beta$) are colour and Dirac indices, respectively, and we remind that $t$ represents the time distance from the source. The stochastic source $\xi_r$ is manifestly zero for all $t \neq 0$. Setting $S_\ell^{\vec\theta} \equiv (D_\ell^{\vec\theta})^{-1}$ and
\begin{equation}
  \eta_{r,\ell}^{\vec\theta}(\vec x, t)\ =\ \sum_{\vec x^\prime} 
  S_\ell^{\vec\theta}(\vec x, t; \vec x^\prime, 0) \cdot
  \xi_r(\vec x^\prime , 0)\,,\qquad \ell=u,d
\end{equation}
one can estimate
\begin{equation}
  \label{eq:C2pt_stoc}
  \sum_{\vec x, a, \alpha} \eta_{r,u}^{a, \alpha, {\vec{\theta}}}(\vec x, t) \cdot \left[ \eta_{r,d}^{a,
      \alpha, {\vec{0}}}(\vec x, t) \right]^\star = C^{2\pt}(t, \vec p)\ +\ \textrm{noise}
\end{equation}
owing to $\gamma_5$-hermiticity $\gamma_5 D_u \gamma_5 = D_d^\dagger$
and $\gamma_5^2 = 1$. In Eq.~(\ref{eq:C2pt_stoc}) the pion momentum $\vec{p}$ is given by $\vec p = 2 \pi \vec{\theta} /L$.
At fixed values of $t^\prime$ (the time distance between the sink and the source) the so-called sequential propagator is computed as
 \begin{equation}
   \varphi_{r, \ell, \ell'}^{{\vec\theta}, {\vec\theta}'}(\vec x, t;
   t')\ =\ \sum_{\vec x^\prime} S_{\ell^\prime}^{\vec\theta^\prime}(\vec x, t; \vec x^\prime, t^\prime)
   \cdot\left[\gamma_5\,\eta_{r,\ell}^{\vec\theta}(\vec x^\prime , t^\prime) \right]\,. 
 \end{equation}
Then, one estimates the three-point function in the Breit frame kinematics from
\begin{equation}
  \sum_{\vec x, a, \alpha, \alpha^\prime} \eta_{r,u}^{a, \alpha, {\vec\theta}}(\vec x, t) 
  \cdot \left[ \varphi_{r, d, u}^{a, \alpha^\prime, {\vec{0}, -\vec\theta}}(\vec x, t;
  t') \right]^\star (\gamma_5 \gamma_0)_{\alpha^\prime \alpha} \propto\ C^{3\pt}_0(t, t', \vec p)\ +\ \textrm{noise}\,.
\end{equation}

We determine $F_\pi(Q^2)$ from the ratio defined in \eq{eq:ratio} in two
different ways: for the first one we compute the double ratio
\begin{equation}
  \label{eq:Mn}
  M_n(t, t', \vec p)\ \equiv\ \frac{R(t, t', \vec p)}{R(t, t', \vec 0)}
\end{equation}
and we extract the pion form factor from its large time distance behavior
\begin{equation}
 \label{eq:numerical}
 F_\pi(Q^2) = \lim_{\substack{t\to\infty \\ (t'-t)\to\infty \\ T\to\infty}}
  M_n(t, t', \vec p)\,.
\end{equation}
We denote this estimate as the \emph{numerical} one. The second estimate consists
in replacing the pseudoscalar two-point function by its analytical
expression, i.e.~we fit
\begin{equation*} 
  C^{2\pt}(t,\vec{p})=\frac{G_\pi^2}{2E_\pi(\vec{p})} \left[
    e^{-E_\pi(\vec{p})t} + e^{-E_\pi(\vec{p})(T-t)} \right] \, ,
\end{equation*}
to the data for the two-point function at large Euclidean times to
determine the amplitude $G_\pi$ and the
energy $E(\vec p)$. Next we define
\begin{equation}
  R_a(t, t', \vec p)\ =\ \frac{2 E_\pi(\vec{p})}{G_\pi^2} \frac{C_0^{3\pt}(t, t', \vec p)}
    {e^{-E_\pi(\vec{p}) t} + e^{-E_\pi(\vec{p}) (T-t)}} \, ,
  \label{eq:Ra}
\end{equation}
where we replace the data for the two-point function by its analytical
expression using the best fit parameters.
Then we calculate the double ratio
\begin{equation}
  \label{eq:Ma}
  M_a(t, t', \vec p)\ \equiv \ \frac{R_a(t, t', \vec{p})}{R_a(t, t', \vec {0})}\,,
\end{equation}
from which the pion form factor can be obtained as
\begin{equation}
    \label{eq:analytical}
    F_\pi(Q^2) = \lim_{\substack{t\to\infty \\ (t'-t)\to\infty \\ T\to\infty}} M_a(t, t', \vec p) ~ .
\end{equation}
The {\em analytical} estimate (\ref{eq:analytical}) may have the advantage of being less noisy than the {\em numerical} one (\ref{eq:numerical}), because the data for the two-point function at large $t$ can
be noisy, in particular for the largest values of $|\vec p|$. 

\begin{figure}[htb!]
\centering
\includegraphics[width=1.0\linewidth]{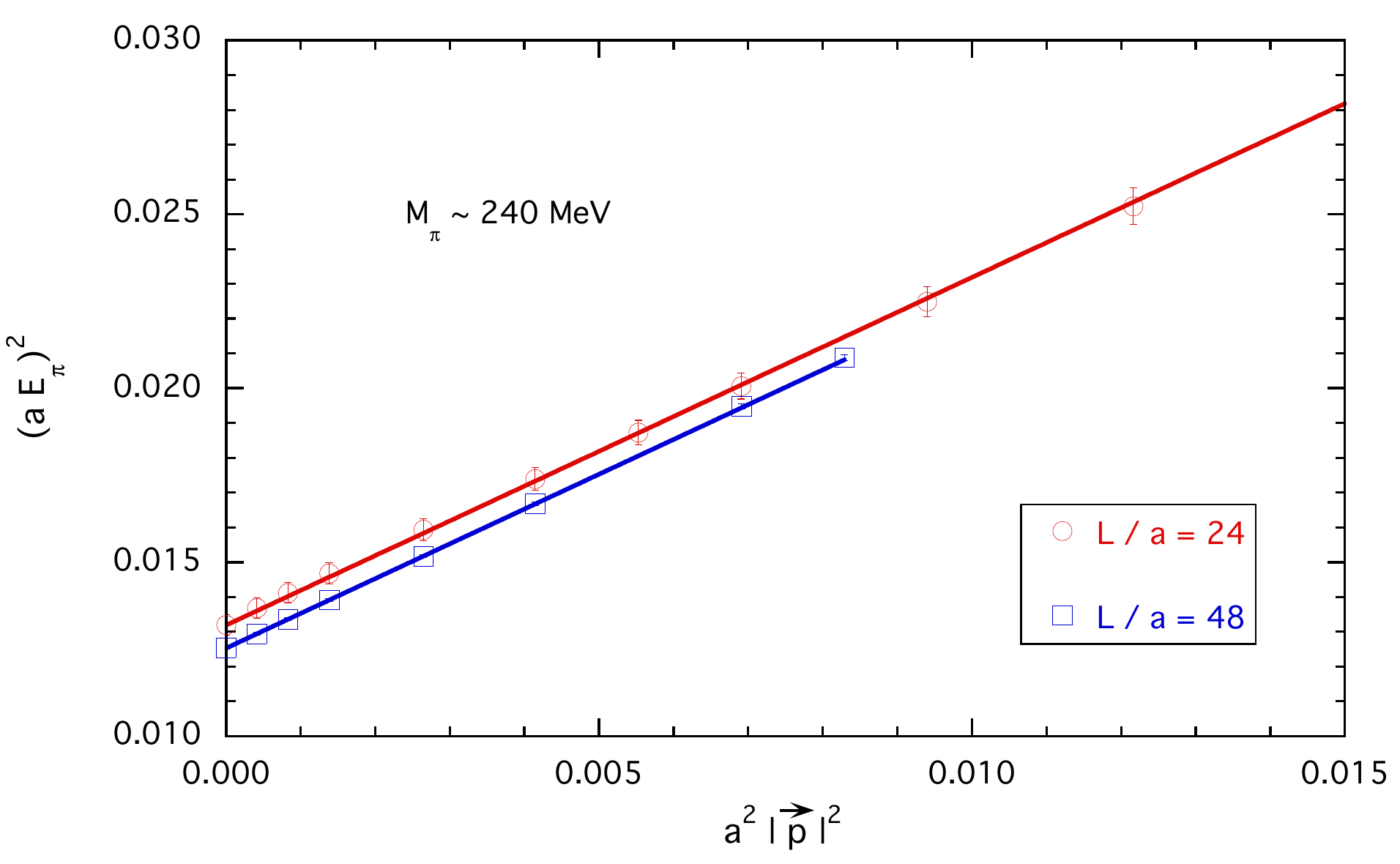}
\vspace{-1.0cm}
\caption{\it \footnotesize Pion dispersion relation for the gauge ensembles cA2.30.24 and cA2.30.48 at a pion mass $M_\pi \simeq 240$ MeV with $L / a = 24$ and $48$, respectively. The solid lines represent the continuum dispersion relation (\ref{eq:Epi}) for the two gauge ensembles.}
\label{fig:pidisp}
\end{figure}

A further improvement is to replace in \eq{eq:Ra} the pion energy $E_\pi(\vec{p})$, extracted from the 2-point correlator $C^{2\pt}(t,\vec{p})$, with the corresponding value from the dispersion relation 
 \begin{equation}
     E_\pi^{disp}\left( \vec{p} = \frac{2 \pi \vec{\theta}}{L} \right) = \sqrt{M_\pi^2(L) + \left( \frac{2 \pi \vec{\theta}}{L} \right)^2} ~ ,
     \label{eq:Epi}
 \end{equation}
where $M_\pi(L)$ is pion mass extracted from the 2-point correlator at rest.
Indeed, in \fig{fig:pidisp} we show the measured energy levels $(aE_\pi)^2$ in
lattice units as a function of the squared momentum $|a \vec{p}|^2$ for the two ensembles cA2.30.24 and c.A2.30.48 with $L / a = 24$ and $48$, respectively. 
The data is described reasonably by the dispersion relation (\ref{eq:Epi}), indicated by the solid lines, up to the largest values of momenta $|a \vec{p}|^2 \sim 0.01$ adopted in this work.
This suggests that the main bulk of finite volume effects (FVEs) on the pion energy $E_\pi(\vec{p})$ originates from those of the pion mass $M_\pi(L)$.

However, the use of non-periodic BCs is expected to produce further FVEs in the dispersion relation (\ref{eq:Epi}). 
Such corrections have been investigated in Ref.~\cite{Jiang:2006gna} using partially quenched ChPT at NLO, finding that the pion momentum $\vec{p} = 2 \pi \vec{\theta} / L$ acquires an additive correction term $2 \pi \vec{K} / L$, namely
 \begin{equation}
     E_\pi( \vec{p}) = \sqrt{M_\pi^2(L) + \left( \frac{2 \pi \vec{K}}{L} + \frac{2 \pi \vec{\theta}}{L} \right)^2} ~ ,
     \label{eq:Epi_chiral}
 \end{equation}
where the components of the vector $\vec K$ are given by
 \begin{equation}
    K_i = - \frac{1}{2 \pi^{3/2} (f_\pi L)^2} \int_0^\infty d\tau ~ \frac{1}{\sqrt{\tau} } ~ e^{-\tau \left(\frac{M_\pi L}{2 \pi} \right)^2} ~ 
             \overline{\Theta}(\tau, \theta_i) ~ \prod_{j \neq i, j=1}^3 \Theta(\tau, \theta_j) 
    \label{eq:kappa}
 \end{equation}
with $\Theta(\tau, \theta) \equiv \sum_{n = - \infty}^\infty e^{-\tau (n + \theta)^2}$ and $\overline{\Theta}(\tau, \theta) \equiv \sum_{n = - \infty}^\infty (n + \theta) e^{-\tau (n + \theta)^2}$ being the elliptic Jacobi function and its derivative.

\begin{figure}[htb!]
\centering
\includegraphics[width=1.0\linewidth]{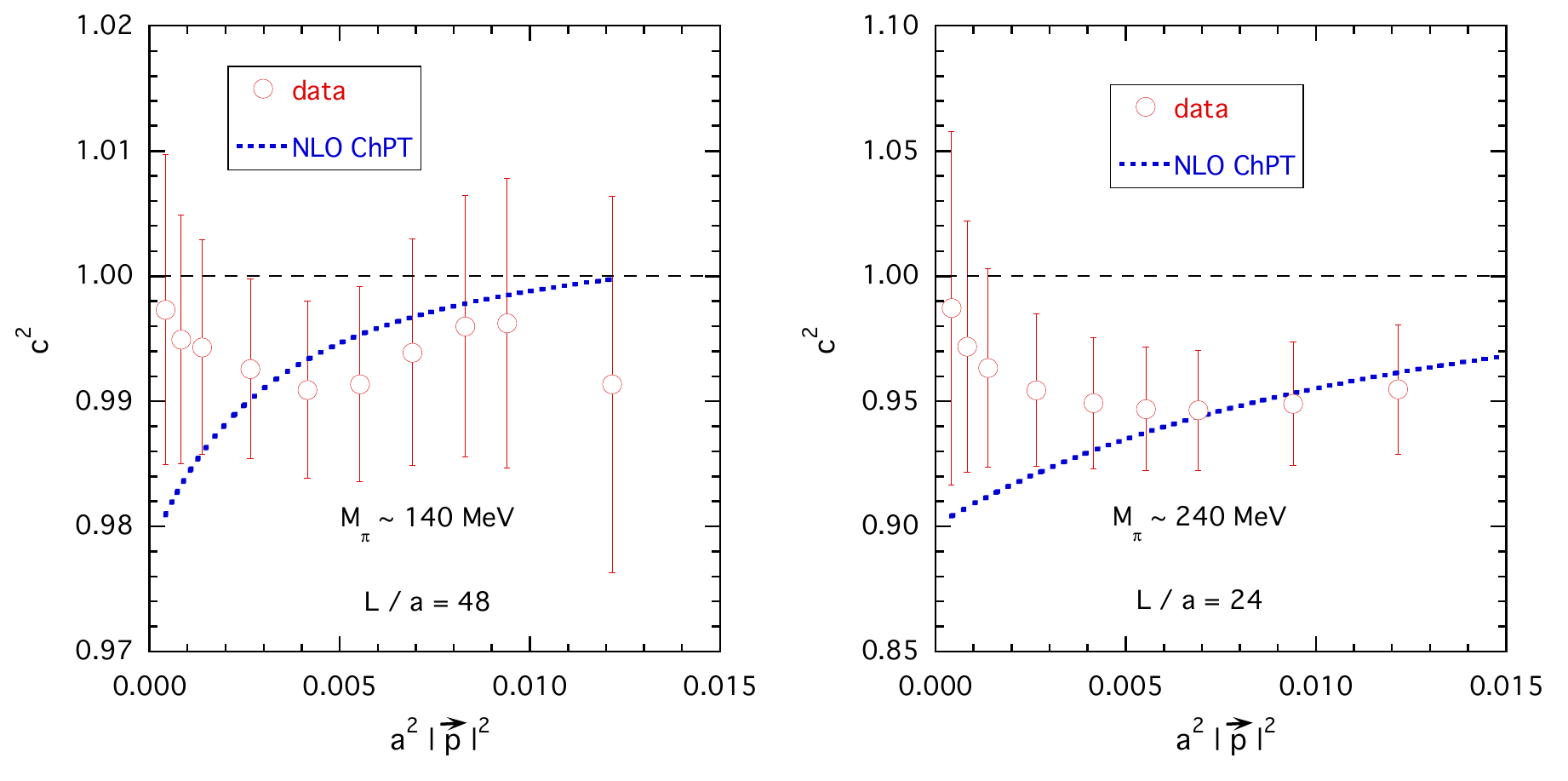}
\vspace{-1.0cm}
\caption{\it \footnotesize The quantity $c^2 = (E_\pi^2(\vec{p}) - M_\pi^2(L)) / |\vec{p}|^2$ versus $a^2 |\vec{p}|^2$ for the gauge ensembles cA2.09.48 (left panel) and cA2.30.24 (right panel). The dashed lines are the predictions of NLO ChPT \cite{Jiang:2006gna} obtained from Eqs.~(\ref{eq:Epi_chiral}-\ref{eq:kappa}). Note the different range of values for $c^2$ in the left and right panels.}
\label{fig:c2}
\end{figure}

For a better visualization of the effects of the additive correction (\ref{eq:kappa}) we consider the dimensionless quantity $c^2$, defined as
 \begin{equation}
     c^2 \equiv \frac{E_\pi^2(\vec{p}) - M_\pi^2(L)}{|\vec{p}|^2} = \frac{|\vec{K} + \vec{\theta}|^2}{|\vec{\theta}|^2} ~ ,
 \end{equation}
which in absence of FVEs on the momentum should be equal to unity.
In \fig{fig:c2} the values of $c^2$ corresponding to the energy $E_\pi(\vec{p})$ and the mass $M_\pi(L)$, extracted from the appropriate 2-point correlators, are shown for various values of $|a \vec{p}|^2$ for the gauge ensemble cA2.09.48 and cA2.30.24.
It can be seen that $c^2$ deviates from unity and its momentum dependence is consistent with the NLO ChPT prediction corresponding to Eqs.~(\ref{eq:Epi_chiral}-\ref{eq:kappa}) at the largest values of $|a \vec{p}|^2$, while the trend of the data is not reproduced at small values of the momentum, even if the present precision does not allow to draw definite conclusions.
This issue certainly deserves further investigations, which are however outside the scope of the present work.  

In order to minimize excited state effects, the source-sink separation is fixed to $t' = T/2$.
On each gauge configuration, multiple source time slices are chosen randomly across the whole time extent, which has been shown to decorrelate measurements from different gauge configurations.
The statistical analysis is performed using the blocked bootstrap method.

\begin{figure}[htb!]
\centering
\includegraphics[width=1.0\linewidth]{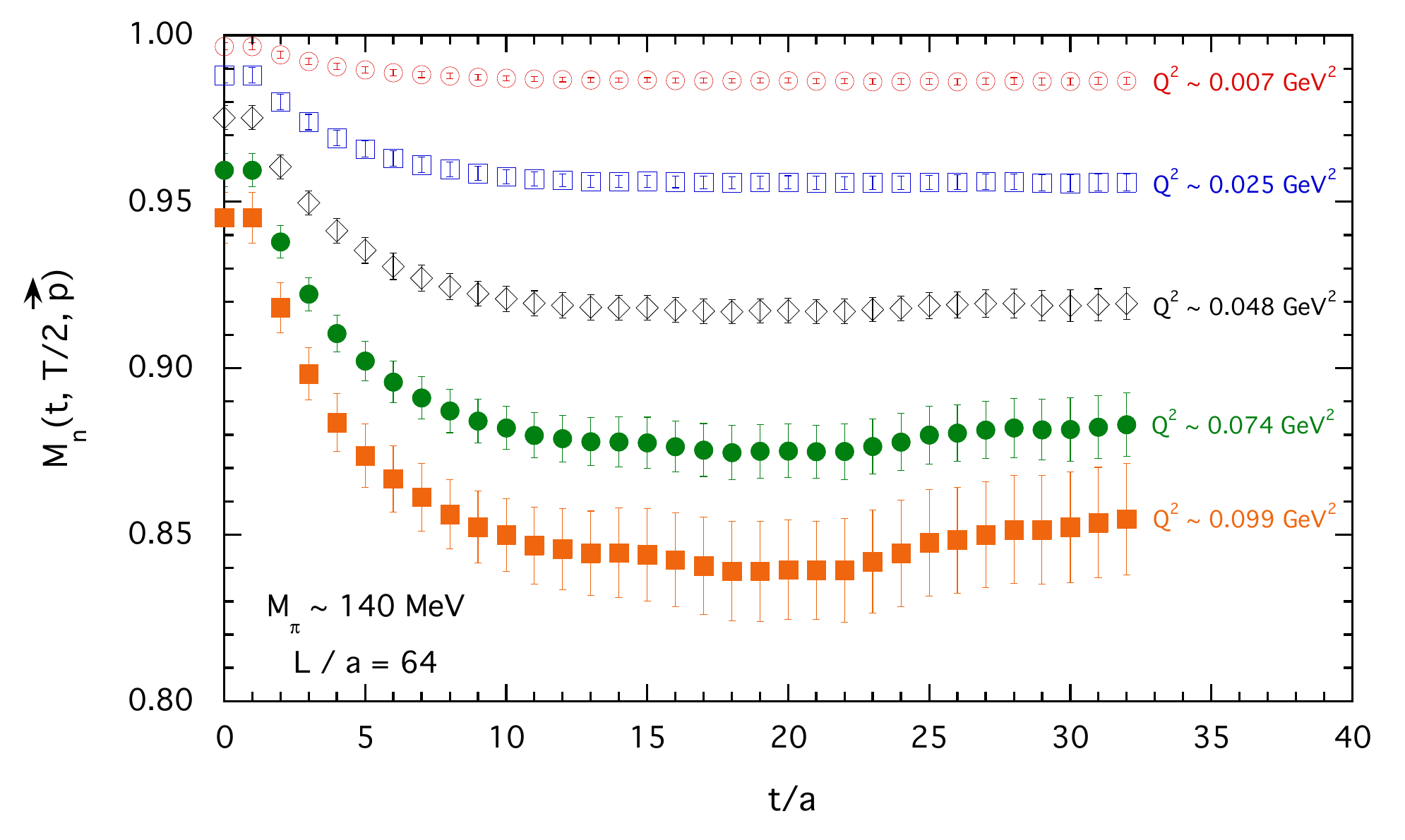}
\vspace{-1.0cm}
\caption{\it \footnotesize The ratio $M_n(t, T/2, \vec{p})$ (see Eq.~(\ref{eq:Mn}) for $t^\prime = T/2$) in the case of the ensemble cA2.09.64 (i.e., $T/2 = 64 a$) for few selected values of the squared 4-momentum transfer $Q^2$. The values of the pion form factor $F_\pi(Q^2)$ are determined from the plateaux corresponding to the time range $[t_{min}, T/2 - t_{min}]$ with $t_{min} = 12 a$. Since the ratio $M_n(t, T/2, \vec{p})$ is symmetric with respect to $T/4$, the plot is limited up to $t = T /4$.}
\label{fig:plateaux}
\end{figure}

Since $M(t, T/2, \vec\theta) = M(T-t, T/2, \vec\theta)$,
we perform the averaging of forward and backward three-point correlation
functions
\[
\overline{M}(t, T/2, \vec\theta) \ = \ \frac{1}{2}[M(t, T/2, \vec\theta) +
  M(T-t, T/2, \vec\theta)]\,. 
\]
The vector form factor $F_\pi(Q^2)$ can then be extracted from the ratio $\overline{M}(t, T/2, \vec\theta)$ for values of $t$ in the range $[t_{min}, T/2 - t_{min}]$, where $t_{min}$ is the time distance at which excited states have decayed sufficiently from both the source and the sink.
The ratio $\overline{M}(t, T/2, \vec\theta)$ is also symmetric with respect to $t = T/4$.
The quality of the plateaux is illustrated in Fig.~\ref{fig:plateaux} for a few selected values of $Q^2$ in the case of the gauge ensemble cA2.09.64.

Before closing this Section, we address briefly the estimate of the renormalization constant  of the vector current, $Z_V$, which can be obtained form the plateau of the ratio (\ref{eq:ZV}).
We remind that the latter one involves 2- and 3-point correlation functions with pion at rest and corresponds to fix the absolute normalization of the pion form factor, $F_\pi(Q^2 = 0) = 1$.
The data for the ratio (\ref{eq:ZV}) exhibit nice plateaux in an extended time region ($t / a \gtrsim 5$) and allow to extract $Z_V$ with a very high statistical precision ($\approx 0.01 \%$).
The resulting values of $Z_V$ do depend upon the quark mass as a pure discretization effect (see also Ref.~\cite{Frezzotti:2008dr}). 
The extrapolation to the chiral limit provides therefore the value of the renormalization constant $Z_V$, which is indeed defined in such a limit.
Using a linear fit in the (bare) quark mass\footnote{A linear dependence on the quark mass is not in contradiction with the ${\cal{O}}(a)$ improvement of the ratio (\ref{eq:ZV}), since terms proportional to $a^2 \mu \Lambda_{QCD}$ may be dominant with respect to terms proportional to $a^2 \mu^2$.} we get $Z_V = 0.6679 ~ (1)_{stat} ~ (1)_{syst}$ at $\beta = 2.10$, where the systematic error corresponds to the uncertainty due to different choices of the time extension of the plateau region in Eq.~(\ref{eq:ZV}).

%%%%%%%%%%%%%%%%%%%%%%%%%%%%%%%%%%%%%%%%%%%%%%%%
\section{Lattice data}
\label{sec:results}

\subsection{Choice of timeslice sources per gauge configuration}

As mentioned in section~\ref{sec:Fpi} we use stochastic timeslice
sources for estimating the pion form factor. Therefore, it is interesting to
investigate how many timeslice sources per gauge configuration are
optimal in order to keep the total statistical error still scaling
like $1/\sqrt{N_\eta}$ with $N_\eta$ being the number of sources per gauge
configuration. Due to correlation between timeslices one expects that too
large values of $N_\eta$ do not improve the final error estimate further. 

\begin{figure}[htb!]
  \centering
  \subfigure{\includegraphics[width=.48\linewidth]{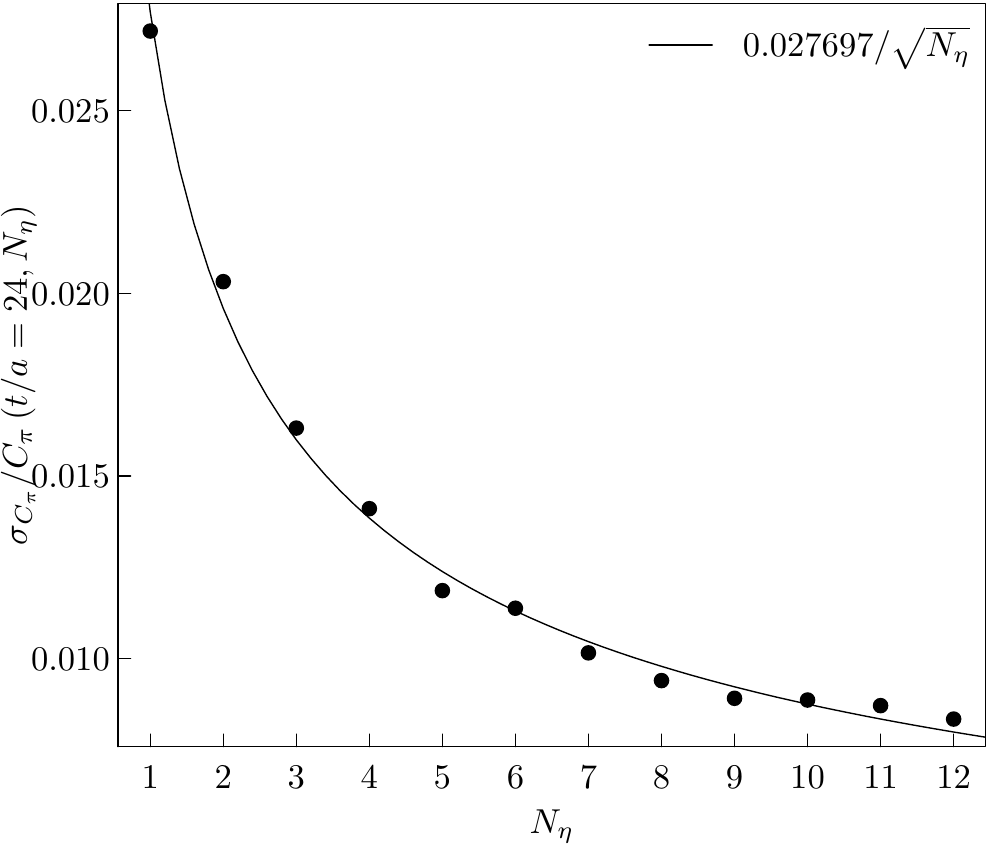}}
  \subfigure{\includegraphics[width=.48\linewidth]{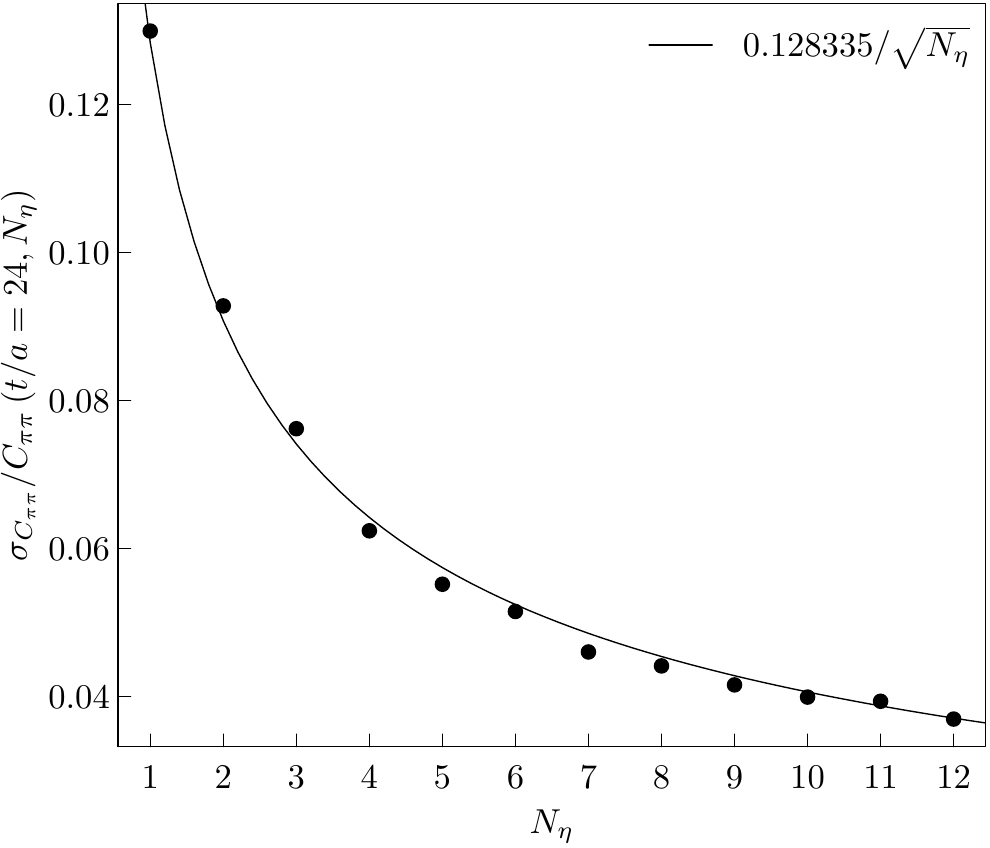}}
  \vspace{-0.25cm}
  \caption{\it \footnotesize Relative error in the pion two point (left) and three point
    (right) functions as a functions
    of the number of sources $N_\eta$ per gauge configuration for
    ensembles cA2.09.48. The solid line represents a fit of
    $c_1/\sqrt{N_\eta}$ to the data points.}
  \label{fig:errorscaling}
\end{figure}

In \fig{fig:errorscaling} we show the relative error of the two-point
(left panel) and the three-point (right panel) correlation functions as a function
of $N_\eta$, at $t / a = 24$ for ensemble cA2.09.48. The different
source times $t_0$ are chosen to be distributed uniformly in the range $0$ to
$T/a - 1$. The solid line represents a fit of the expected
$\sqrt{N_\eta}^{-1}$ behaviour to the data.

We observe that the error follows the $\sqrt{N_\eta}^{-1}$ behaviour
basically up to $N_\eta=12$, where we stopped. Since from $N_\eta=8$
on, the error does not improve significantly anymore, we fix
$N_\eta^{T=96}=12$. This amounts to a mean distance of $12$
between source timeslices. We keep this mean difference fixed also for
the other lattice volumes, i.e. $N_\eta^{T=128}=16$, $N_\eta^{T=64}=8$
and $N_\eta^{T=48}=4$.

%%%%%%%%%%%%%%%%%%%%%%%%%%%%%%%%%%%%%%%%%%%%%%%%
\subsection{Pion Electromagnetic Form Factor}
 
The lattice data obtained for $F_\pi(Q^2)$ as a function of $Q^2$ in physical units for all the gauge ensembles of Table \ref{tab:setup} is shown in \fig{fig:fpiraw} and collected in the Appendix together with the values chosen for the pion momentum. 
In the left panel data is shown up to $Q^2 = 0.250\ \mathrm{GeV}^2$,
while the right panel restricts $Q^2$ to values smaller than
$0.12\ \mathrm{GeV}^2$. In addition to our lattice data we also show
experimental data from CERN~\cite{Amendolia:1986wj}. 

\begin{figure}[htb!]
\centering
\subfigure{\includegraphics[width=0.48\linewidth]{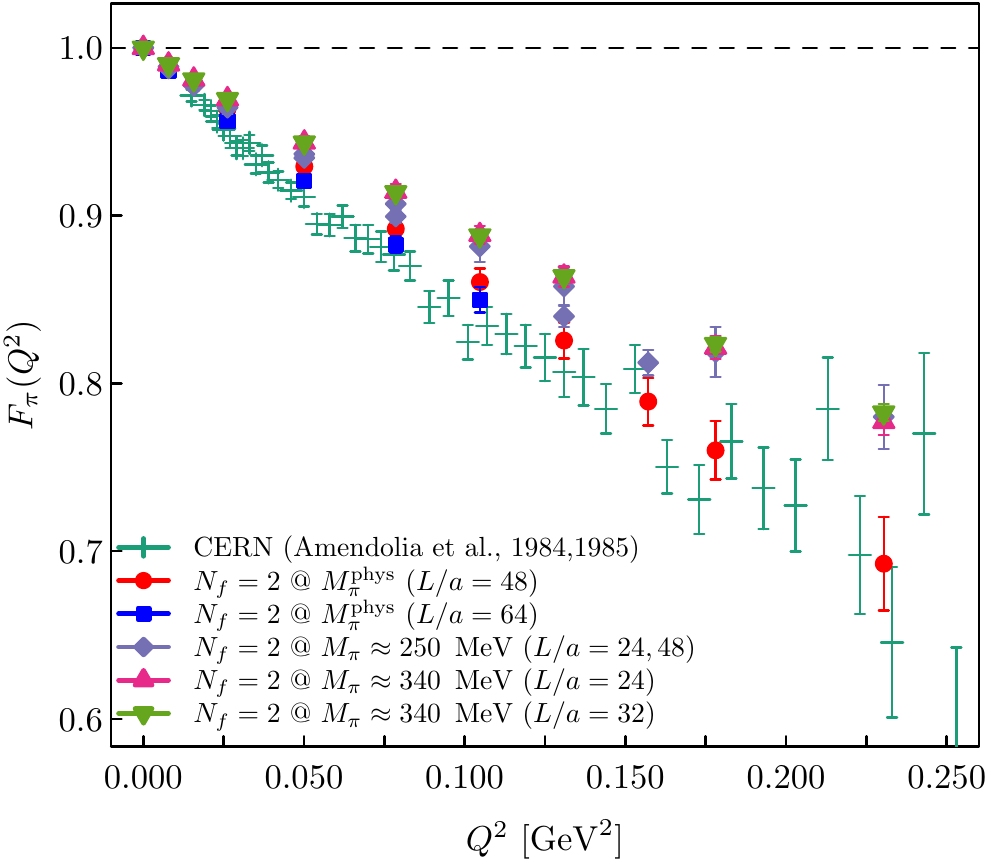}}
\subfigure{\includegraphics[width=0.48\linewidth]{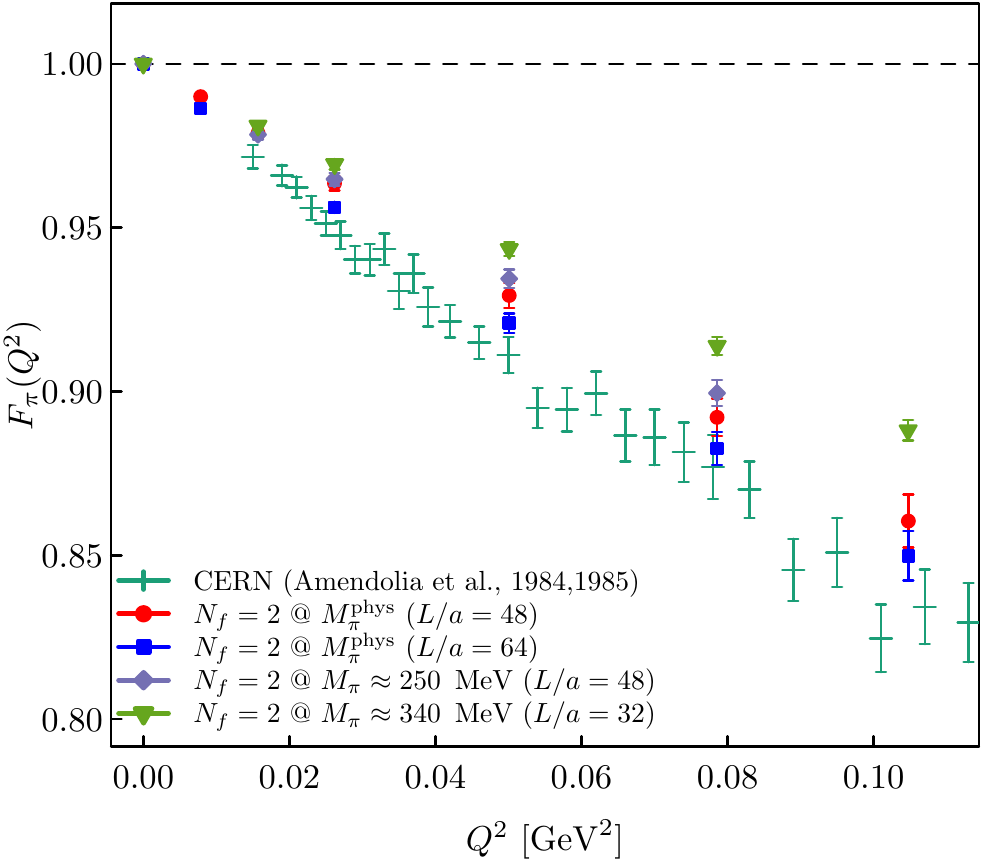}}
\vspace{-0.25cm}
\caption{\it \footnotesize Data for the vector form factor $F_\pi$ as a function of $Q^2$ for all ensembles used in this work. In addition we show experimental results from CERN~\cite{Amendolia:1986wj}. The right panel is a restriction of the left panel to values of $Q^2 < 0.12 ~ \mathrm{GeV}^2$.}
\label{fig:fpiraw}
\end{figure}

It is visible that the errors of our lattice data are compatible with the ones of the
experimental data. In particular for small $Q^2$ (right panel) the errors of the
lattice data are significantly smaller than the errors of the
experimental single data points. Of course, the experimental points
have a much denser coverage of $Q^2$ values. However, thanks to
non-periodic boundary conditions our lattice data covers $Q^2$ values below
the range where experimental data is available.
Moreover, our lattice data at
the physical pion point and at the largest volume is compatible with the
experimental data within statistical errors.

\begin{figure}[htb!]
\centering
\includegraphics[width=1.0\linewidth]{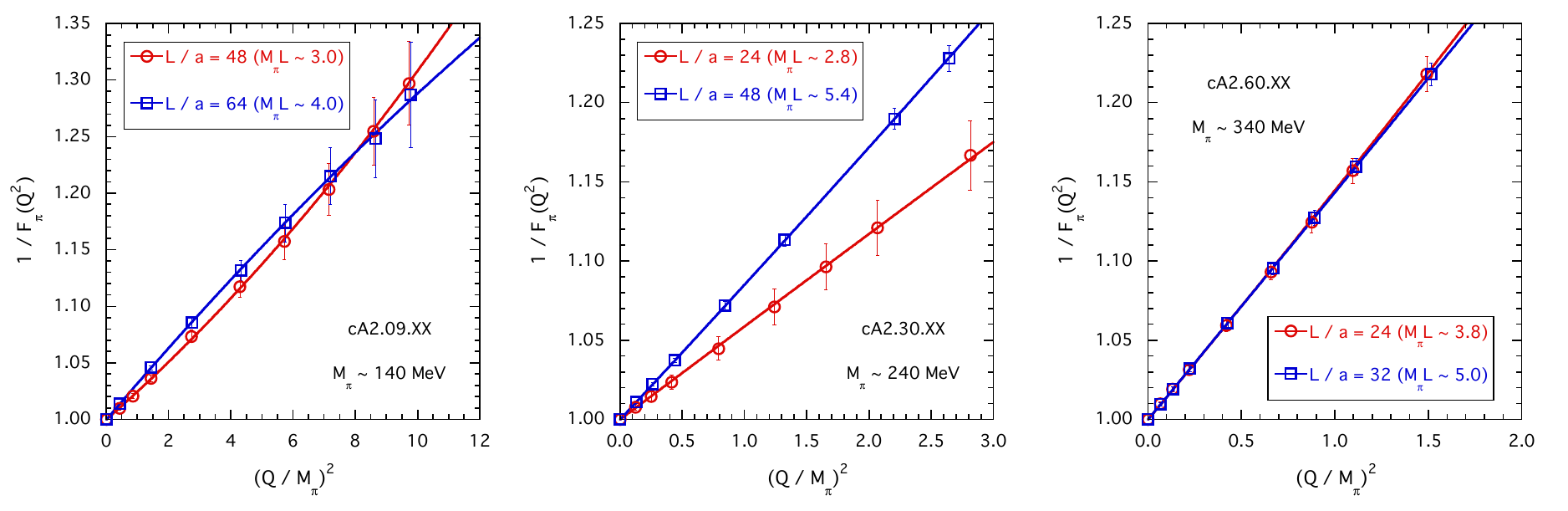}
\vspace{-1.0cm}
\caption{\it \footnotesize Data for the inverse pion form factor $1 / F_\pi$ as a function of $(Q / M_\pi)^2$ for the six ensembles used in this work. In the insets the values of $M_\pi L$ are shown. The solid lines represent the results of the quadratic fit (\ref{eq:quadratic}).}
\label{fig:FFpion}
\end{figure}

We have collected in Fig.~\ref{fig:FFpion} the lattice data for the inverse pion form factor $1 / F_\pi$ versus the dimensionless variable $(Q/ M_\pi)^2$ for the six gauge ensembles of Table \ref{tab:setup}.
It can be seen that the data for $1 / F_\pi$ exhibits an almost linear behavior with $Q^2$, as expected from VMD arguments.
Actually the solid lines in Fig.~\ref{fig:FFpion} represent the results of a quadratic fit in $Q^2$,
 \begin{equation}
  \frac{1}{F_\pi(Q^2)} = 1 + s_\pi^\prime \left( \frac{Q}{M_\pi} \right)^2 + c_\pi^\prime \left( \frac{Q}{M_\pi} \right)^4 ~ ,
  \label{eq:quadratic}
 \end{equation}
where we find that $c_\pi^\prime \ll (s_\pi^\prime)^2$ in accord with the VMD hypothesis.
Moreover, for each pion mass the data for two different lattice volumes are compared in Fig.~\ref{fig:FFpion}. It can clearly be seen that finite volume effects are relevant for $M_\pi L \lesssim 3$.

%%%%%%%%%%%%%%%%%%%%%%%%%%%%%%%%%%%%%%%%%%%%%%%%
\section{Chiral Extrapolation and Finite Volume Effects}
\label{sec:method}

%%%%%%%%%%%%%%%%%%%%%%%%%%%%%%%%%%%%%%%%%%%%%%%%
\subsection{Chiral Extrapolation}
\label{sec:chiral}

Within SU(2) Chiral Perturbation Theory (ChPT) the expansion of the pion form factor $F_\pi(Q^2)$ in powers of the squared pion mass reads as
\begin{equation}
      F_\pi(Q^2) = 1 + \Delta F_\pi^{\rm NLO}(Q^2) + \Delta F_\pi^{\rm NNLO}(Q^2) + ... ~ ,
     \label{eq:ChPT}
\end{equation}
where $\Delta F_\pi^{\rm NLO}(Q^2)$ is the next-to-leading order (NLO) term and $\Delta F_\pi^{\rm NNLO}(Q^2)$ the NNLO one.
Both are known~\cite{Gasser:1983yg,Bijnens:1998fm} and the NLO term is explicitly given by
\begin{equation}
  \label{eq:NLO}
  \Delta F_\pi^{\rm NLO}(Q^2) = - \frac{\xi}{3} \frac{Q^2}{M_\pi^2} \left[ \overline{\ell}_6 - \log\frac{\xi}{\xi^{phys}} - 1 + 
                                             R\left(\frac{Q^2}{M_\pi^2}\right) \right] ~ ,
\end{equation}
where $\overline{\ell}_6$ is an SU(2) LEC, $\xi \equiv M_\pi^2 / (4 \pi f_\pi)^2$ and
\begin{equation}
  R(w) = \frac{2}{3} + \left( 1+ \frac{4}{w} \right) \left[ 2 + \sqrt{1 +\frac{4}{w}}
               \log\frac{\sqrt{1 + \frac{4}{w}} - 1}{\sqrt{1 + \frac{4}{w}} + 1} \right] ~ .
\end{equation}

Let's define the slope $s_\pi$ and the curvature $c_\pi$ of the pion form factor in terms of its expansion in powers of $(Q / M_\pi)^2$ as
 \begin{equation}
    F_\pi(Q^2) = 1 - s_\pi \frac{Q^2}{M_\pi^2} + c_\pi \frac{Q^4}{M_\pi^4} + 
                         {\cal{O}}\left( \frac{Q^6}{M_\pi^6} \right) ~ .
 \end{equation}
 At NLO one has
  \begin{eqnarray}
      \label{eq:slope_NLO}
      s_\pi^{\rm NLO} & = & \frac{1}{3} \xi \left[ \overline{\ell}_6 - \log\frac{\xi}{\xi^{phys}} - 1 \right] ~ , \\
      \label{eq:curvature_NLO}
      c_\pi^{\rm NLO} & = & \frac{1}{30} \xi ~ ,
  \end{eqnarray}
which show that the LEC $\overline{\ell}_6$ governs only the value of the slope $s_\pi^{\rm NLO}$.
Once the value of $\overline{\ell}_6$ is fixed by the reproduction of the experimental value of the pion charge radius, i.e.~$\overline{\ell}_6 \simeq 14.6$ (see Ref.~\cite{Frezzotti:2008dr}), it turns out that $c_\pi^{\rm NLO} \ll (s_\pi^{\rm NLO})^2$ , which is in contradiction with the VMD phenomenology observed both in the experimental data and  in our lattice results up to a pion mass of $\simeq 340$ MeV (see Fig.~\ref{fig:FFpion}).
In Ref.~\cite{Frezzotti:2008dr} it was found that, using $\overline{\ell}_6 \simeq 14.6$ the NLO term (\ref{eq:NLO}) works only for very low values of both $Q^2$ ($Q^2 \lesssim 0.03$ GeV$^2$) and the pion mass ($M_\pi \lesssim 300$ MeV).
Effects from NNLO and higher order terms in the chiral expansion (\ref{eq:ChPT}) become more and more important as the value of $Q^2$ increases.
In particular, the curvature $c_\pi$ is found to be almost totally dominated by NNLO effects \cite{Frezzotti:2008dr}. The latter however depend on several LECs (see Ref.~\cite{Bijnens:1998fm}).

In order to avoid the need of many LECs let's consider the inverse of the pion form factor.
Using Eq.~(\ref{eq:ChPT}) the SU(2) ChPT expansion of $1 / F_\pi(Q^2)$ reads as
\begin{equation}
     \frac{1}{F_\pi(Q^2)} = 1 - \Delta F_\pi^{\rm NLO}(Q^2) + \left[ \left( \Delta F_\pi^{\rm NLO}(Q^2) \right)^2 - 
                                        \Delta F_\pi^{\rm NNLO}(Q^2) \right]+ ... ~ ,
     \label{eq:ChPT_inv}
\end{equation}
where on the r.h.s.~the term in the square brackets represent the NNLO correction.
Because of the observed VMD phenomenology (see Fig.~\ref{fig:FFpion}), the NNLO term $\Delta F_\pi^{\rm NNLO}(Q^2)$ in Eq.~(\ref{eq:ChPT_inv}) is expected to be almost compensated by the square of the NLO one $\Delta F_\pi^{\rm NLO}(Q^2)$, leading to a small residual NNLO correction in the inverse pion form factor.
This means that $1 / F_\pi(Q^2)$ is dominated by the NLO approximation at least in the range of values of $Q^2$ and $M_\pi$ covered by our simulations, i.e.~$Q^2 \lesssim 0.25$ GeV$^2$ and $M_\pi \lesssim 340$ MeV.
Thus, we can profit from the above feature by using the following ansatz for the chiral extrapolation of the inverse pion form factor 
\begin{equation}
   \label{eq:chiral}
   \frac{1}{F_\pi(Q^2)} = 1 + \frac{\xi}{3} \frac{Q^2}{M_\pi^2} \left[\overline{\ell}_6 - \log\frac{\xi}{\xi^{phys}} - 1 + 
       R\left(\frac{Q^2}{M_\pi^2}\right) \right] + \frac{\xi^2}{6}\frac{Q^2}{M_\pi^2} \left[b_1 + b_2 \frac{Q^2}{M_\pi^2} \right] ~ ,
\end{equation}
where the last term in the r.h.s.~parametrizes NNLO effects, which we stress are expected to be small.
Eq.~(\ref{eq:chiral}) depends only on three unknowns, namely $\overline{\ell}_6$, $b_1$ and $b_2$, which we determine by fitting our data.

%%%%%%%%%%%%%%%%%%%%%%%%%%%%%%%%%%%%%%%%%%%%%%%%
\subsection{Finite Volume Effects}
\label{sec:FVE}

As illustrated in Fig.~\ref{fig:FFpion}, our data for the pion form factor suffer from finite volume effects (FVEs). 
In this work we follow three strategies to correct for FVEs, profiting from the two lattice volumes available at each value of the quark mass.

We first introduce the FVE factor $K_{\rm FVE}(Q^2, L)$ defined as
 \begin{equation}
     F_\pi(Q^2, L) = F_\pi(Q^2, \infty) + K_{\rm FVE}(Q^2, L) ~ ,
     \label{eq:KFVE}
 \end{equation}
which implies
 \begin{equation}
     \frac{1}{F_\pi(Q^2, L)} = \frac{1}{F_\pi(Q^2, \infty)} \left[ 1 - \frac{1}{F_\pi(Q^2, \infty)}  
                                            K_{\rm FVE}(Q^2, L) \right] ~ ,
     \label{eq:Fpi_inv}
 \end{equation}
where $1 / F_\pi(Q^2, \infty)$ is given by Eq.~(\ref{eq:chiral}).

The three strategies are as follows:
\begin{itemize}
\item[A)] make use of the SU(2) ChPT prediction derived at NLO in the Breit frame~\cite{Jiang:2008te,Colangelo:2016wgs}. The correction factor $K_{\rm FVE}(Q^2, L)$ reads explicitly
  \begin{eqnarray}
    K_{\rm FVE}(Q^2, L) & = & \frac{C}{f_\pi^2} \left\{ \int_0^1 dx ~ I_{1/2}\left[ (1 - 2x) 
                                          \frac{2\pi \vec{\theta}}{L}; M_\pi^2 + x (1 - x) Q^2 \right] 
                                          \right. \nonumber \\
                                 & - & \left. I_{1/2}\left[ \frac{2\pi \vec{\theta}}{L}; M_\pi^2 \right] \right\} ~ ,
    \label{eq:KChPT}
  \end{eqnarray}
where $C$ is a parameter to be determined in the fitting procedure, $Q^2 = 4 (2 \pi \vec{\theta} / L)^2$ and
 \begin{equation}
    I_{1/2}\left[ \frac{2\pi \vec{\theta}}{L}; M_\pi^2 \right] = \frac{1}{2 \pi^{3/2} L^2} 
    \int_0^\infty d\tau ~ \frac{1}{\sqrt{\tau} } ~ e^{-\tau \left(\frac{M_\pi L}{2 \pi} \right)^2} ~ 
    \left[ \prod_{i = 1}^3 \Theta(\tau, \theta_i) - \left( \frac{\pi}{\tau} \right)^{3/2} \right]
    \label{eq:I1/2}
 \end{equation}
with $\Theta(\tau, \theta) \equiv \sum_{n = - \infty}^\infty e^{-\tau (n + \theta)^2}$ being the elliptic Jacobi function.

\item[B)] use a phenomenological ansatz, inspired by the asymptotic expansion of \eq{eq:KChPT}, given by
 \begin{equation}
     K_{\rm FVE}(Q^2, L) = \frac{Q^2}{M_\pi^2} \left[ C_1 + C_2 \frac{Q^2}{M_\pi^2} \right] \frac{\xi}{(M_\pi L)^{3/2}} 
                                    \cdot e^{-M_\pi L}  ~ ,
 \end{equation}
where $C_1$ and $C_2$ are parameters to be determined in the fitting procedure.

\item[C)] use only the largest volume available at each pion mass and assume that FVEs are negligible for these volumes (i.e.~putting $K_{\rm FVE} = 0$). 

\end{itemize}

We want to point out that our fitting ansatz (\ref{eq:Fpi_inv}) is defined in terms of dimensionless quantities only, namely $\xi$, $Q^2 / M_\pi^2$ and $M_\pi L$, and therefore the knowledge of the lattice scale is not required.

%%%%%%%%%%%%%%%%%%%%%%%%%%%%%%%%%%%%%%%%%%%%%%%%
\section{Extrapolations to the physical point }
\label{sec:extrapolations}

In this Section we perform the chiral and infinite volume extrapolations of the lattice data adopting our fitting ansatz (\ref{eq:Fpi_inv}).
Various sources of systematic effects have been taken into account, namely

\begin{itemize}

\item the {\em numerical} and {\em analytical} estimates of the pion form factor given by Eqs.~(\ref{eq:numerical}) and (\ref{eq:analytical}), respectively. The corresponding uncertainty will be denoted by $()_{ratio}$;

\item the time extensions $[t_{min}, T/2 - t_{min}]$ chosen for the plateaux of the double ratios (\ref{eq:Mn}) and (\ref{eq:Ma}) corresponding to $t_{min} = 10$ and $12$. The corresponding uncertainty will be denoted by $()_{fit-range}$;

\item either the inclusion of all the six gauge ensembles of Table \ref{tab:setup} or the restriction to the two gauge ensembles cA2.09.XX at the physical pion mass. The corresponding uncertainty will be denoted by $()_{M_\pi}$;

\item either the inclusion ($b_1 \neq 0$ and $b_2 \neq 0$) or the exclusion ($b_1 = b_2 = 0$) of the NNLO effects in Eq.~(\ref{eq:chiral}). The corresponding uncertainty will be denoted by $()_{ChPT}$;

\item the FVEs evaluated according to the three procedures A, B and C, described in Section \ref{sec:FVE}. The corresponding uncertainty will be denoted by $()_{FVE}$;

\item the inclusion of all $Q^2$ values or the restriction to $Q^2 \leq 2 M_\pi^2$. The corresponding uncertainty will be denoted by $()_{Q^2-range}$.

\end{itemize}

\begin{figure}[htb!]
\centering
\includegraphics[width=1.0\linewidth]{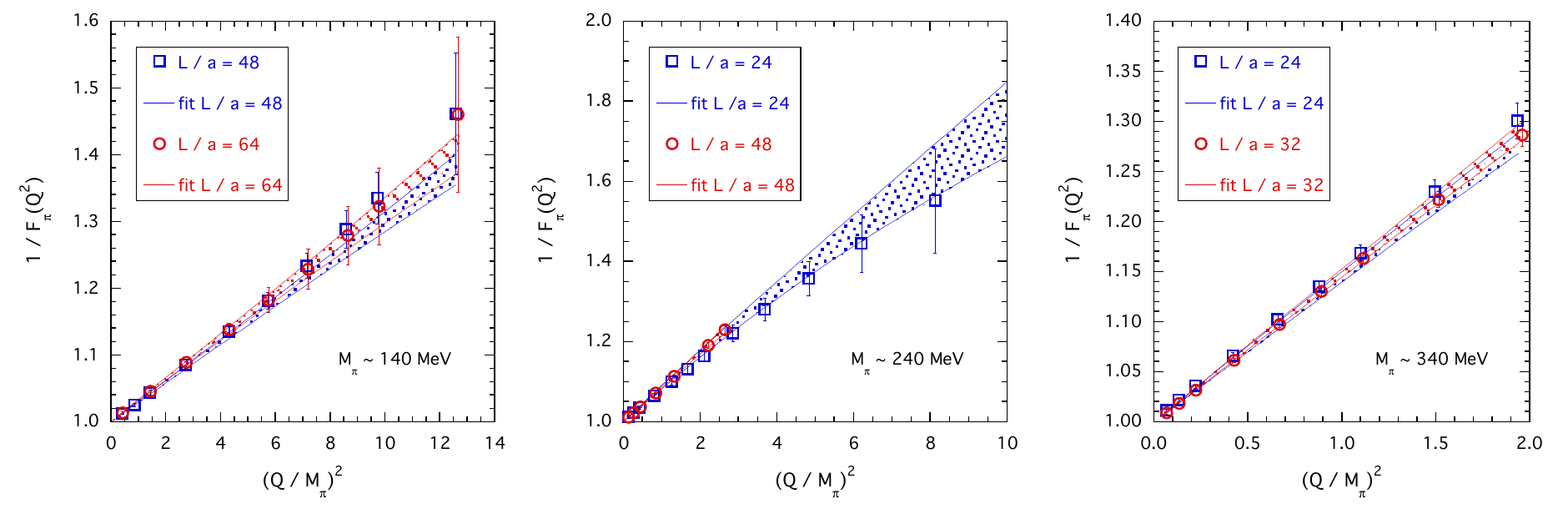}
\vspace{-1.0cm}
\caption{\it \footnotesize Results of our fitting procedure of the inverse pion form factor, based on Eq.~(\ref{eq:Fpi_inv}) with $1 / F_\pi(Q^2, \infty)$ and $K_{\rm FVE}(Q^2, L)$ given by Eqs.~(\ref{eq:chiral}) and (\ref{eq:KChPT}), respectively, for all the six ensembles of Table \ref{tab:setup}. The bands correspond to statistical uncertainties only.}
\label{fig:fits}
\end{figure}

\begin{figure}[htb!]
\centering
\includegraphics[width=0.95\linewidth]{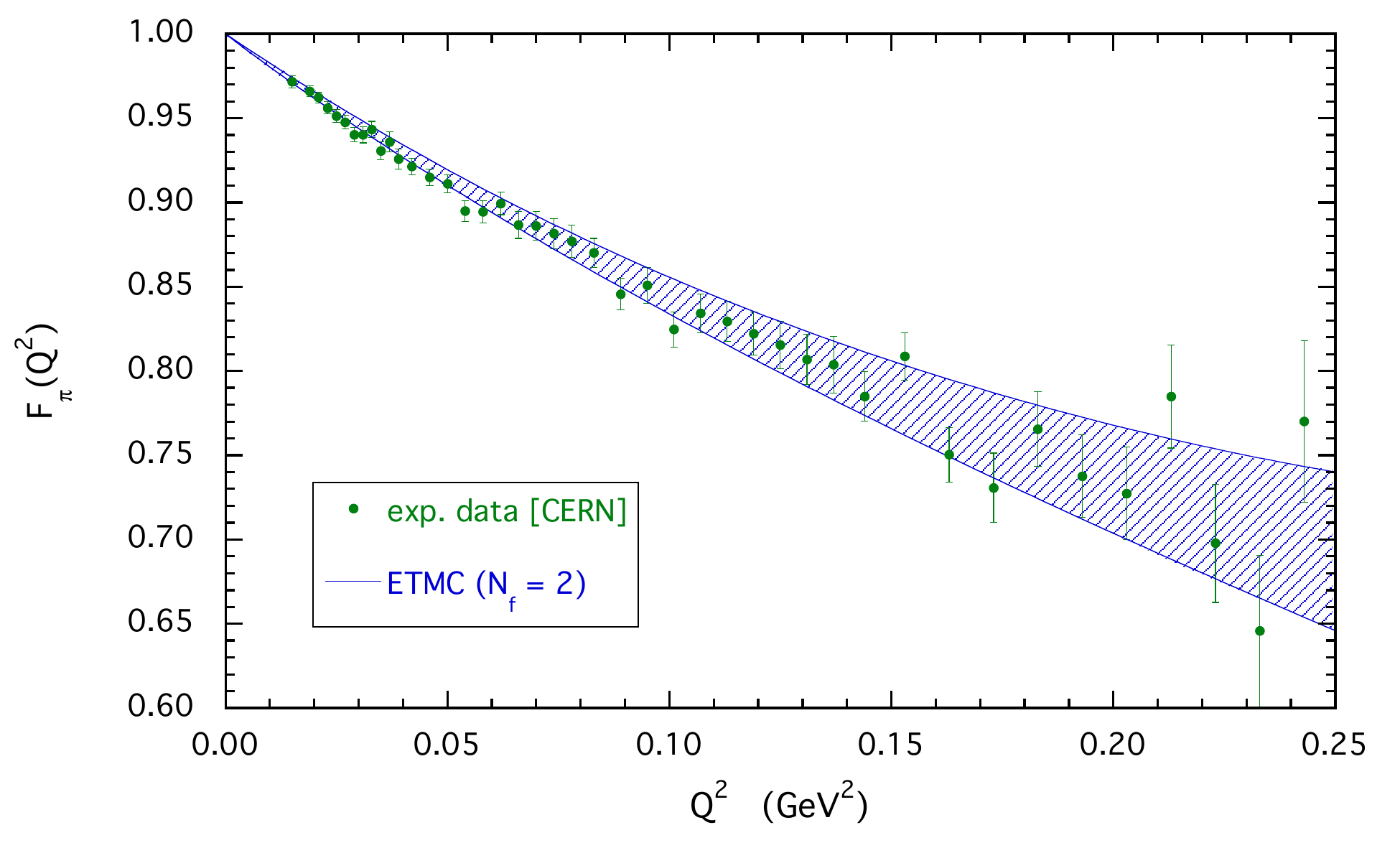}
\vspace{-0.5cm}
\caption{\it \footnotesize Comparison of the $N_f = 2$ ETMC values for $F_\pi(Q^2)$, extrapolated to the physical pion point and to infinite volume limit, with the experimental data from CERN \cite{Amendolia:1986wj}. The band includes both statistical and systematic errors.}
\label{fig:comparison}
\end{figure}

The quality of our fitting procedure is illustrated in Fig.~\ref{fig:fits} for all the six ensembles used in this work\footnote{The values of the fitting parameters are determined by a $\chi^2$-minimization procedure adopting an uncorrelated $\chi^2$. The resulting values of $\chi^2 / \mbox{d.o.f.}$ do not exceed $\simeq 1.3$. The results of the various fits are averaged according to Eq.~(28) of Ref.~\cite{Carrasco:2014cwa}}.
The results for the pion form factor, extrapolated at the physical pion point and in the infinite volume limit, are compared with the experimental data in Fig.~\ref{fig:comparison}.

As far as the pion charge radius is concerned, our fitting ansatz (\ref{eq:Fpi_inv}) implies that at the physical pion point and the infinite volume limit one has
 \begin{equation}
     \langle r^2 \rangle_\pi = \frac{1}{(4 \pi f_\pi)^2} \left[ 2 (\overline{\ell}_6 - 1) + \frac{M_\pi^2}{(4 \pi f_\pi)^2} b_1 \right] ~ .
     \label{eq:slope_NNLO}
 \end{equation}
Thus, our final result at a fixed lattice spacing ($a \simeq 0.09$ fm) reads 
 \begin{eqnarray}
     \langle r^2 \rangle_\pi & = & 0.443 ~ (21)_{stat} ~ (7)_{ratio} ~ (1)_{fit-range} ~ (7)_{M_{\pi}} ~ (6)_{ChPT} ~ (15)_{FVE} ~ 
                                                  (6)_{Q^2-range} ~ \mbox{fm}^2  \nonumber \\
                                         & = & 0.443 ~ (21)_{stat} ~ (20)_{syst} ~ \mbox{fm}^2 \nonumber \\
                                         & = & 0.443 ~ (29) ~ \mbox{fm}^2 ~ ,
    \label{eq:r2pi_final}
 \end{eqnarray}
which is consistent with the experimental value $\langle r^2 \rangle_\pi^{exp.} = 0.452~(11)$ fm$^2$~\cite{Olive:2016xmw}.
This suggests that the impact of discretization effects on our result (\ref{eq:rpi2_final}) could be small with respect to the other sources of uncertainties.

The lattice calculations of $\langle r^2 \rangle_\pi$ have been analyzed recently by FLAG and are collected in Table 22 of Ref.~\cite{Aoki:2016frl}.
Four results satisfy the FLAG quality criteria, namely: $\langle r^2 \rangle_\pi = 0.441~(66)$ fm$^2$~\cite{Brommel:2006ww} ($N_f = 2$), $\langle r^2 \rangle_\pi = 0.456~(38)$ fm$^2$~\cite{Frezzotti:2008dr} ($N_f = 2$), $\langle r^2 \rangle_\pi = 0.481~(35)$ fm$^2$~\cite{Brandt:2013dua} ($N_f = 2$) and $\langle r^2 \rangle_\pi = 0.403~(19)$ fm$^2$~\cite{Koponen:2015tkr} ($N_f = 2 + 1 + 1$).
Our finding (\ref{eq:r2pi_final}) is nicely consistent with all the above lattice results.

The value of the NLO SU(2) LEC $\overline{\ell}_6$, appearing in Eq.~(\ref{eq:slope_NNLO}) and corresponding to our result (\ref{eq:r2pi_final}), is equal to 
 \begin{eqnarray}
     \overline{\ell}_6 & = & 16.21 ~ (76)_{stat} ~ (25)_{ratio} ~ (3)_{fit-range} ~ (26)_{M_{\pi}} ~ (24)_{ChPT} ~ (50)_{FVE} ~ 
                                         (20)_{Q^2-range}  \nonumber \\
                               & = & 16.21 ~ (76)_{stat} ~ (70)_{syst} \nonumber \\
                               & = & 16.21 ~ (1.03) ~ .
    \label{eq:ell6}
 \end{eqnarray}

%%%%%%%%%%%%%%%%%%%%%%%%%%%%%%%%%%%%%%%%%%%%%%%%
\section{Summary and Discussion}
\label{sec:conclusions}

We have presented an investigation of the electromagnetic pion form factor, $F_\pi(Q^2)$, at small values of the four-momentum transfer $Q^2$ ($\lesssim 0.25\gev^2$), based on the gauge configurations generated by ETMC with $N_f = 2$ twisted-mass quarks at maximal twist including a clover term.
Momentum is injected using non-periodic boundary conditions and the calculations are carried out at a fixed lattice spacing ($a \simeq 0.09$ fm) and with pion masses equal to its physical value, 240 MeV and 340 MeV.
We have successfully analyzed our data using Chiral Perturbation Theory at next-to-leading order in the light-quark mass.
For each pion mass two different lattice volumes are used to take care of finite size effects.
Our final result for the squared charge radius is $\langle r^2 \rangle_\pi = 0.443~(29)$ fm$^2$, where the error includes several sources of systematic errors except the uncertainty related to discretization effects. 
The corresponding value of the SU(2) low-energy constant $\overline{\ell}_6$ is equal to $\overline{\ell}_6 = 16.2 ~ (1.0)$.
Our result is consistent with the experimental value $\langle r^2 \rangle_\pi^{exp.} = 0.452~(11)$ fm$^2$~\cite{Olive:2016xmw} as well as with other lattice estimates (see Ref.~\cite{Aoki:2016frl}).
This suggests that the impact of discretization effects on our result could be small with respect to the other sources of uncertainties.

%%%%%%%%%%%%%%%%%%%%%%%%%%%%%%%%%%%%%%%%%%%%%%%%
\begin{acknowledgments}
We thank the members of ETMC for the most enjoyable collaboration. 
The computer time for this project was made available to us by the John von Neumann-Institute for Computing (NIC) on the Jureca and Juqueen systems in J{\"u}lich, and by the Gauss Centre for Supercomputing under project No PR74YO on the GCS Supercomputer SuperMUC at Leibniz Supercomputing Centre. 
This work was granted access also to the HPC resources of CINES and IDRIS under the allocation 52271 made by GENCI. 
This project was funded by the DFG as a project in the Sino-German CRC110 and by the Horizon 2020 research and innovation program of the European Commission under the Marie Sklodowska-Curie grant agreement No 642069. 
S.B.~is supported by the latter program.
The open source software packages tmLQCD~\cite{Jansen:2009xp,Deuzeman:2013xaa,Abdel-Rehim:2013wba}, Lemon~\cite{Deuzeman:2011wz}, DD$\alpha$AMG~\cite{Alexandrou:2016izb} and R~\cite{R:2005} have been used.
\end{acknowledgments}

%%%%%%%%%%%%%%%%%%%%%%%%%%%%%%%%%%%%%%%%%%%%%%%%
\section*{Appendix}

In Tables \ref{tab:table1}-\ref{tab:table6} we collect the values adopted for the vector $\vec{\theta} = \{ \theta, \theta, \theta \}$, the squared 3-momentum $|\vec{p}|^2 = 4 \pi^2 |\vec{\theta}|^2 / L^2$ in lattice units, the squared 4-momentum transfer $Q^2 = - q^2 = 4 |\vec{p}|^2$ in units of the pion mass and the values of the pion form factor $F_\pi(Q^2)$ in the case of the six ensembles of Table \ref{tab:setup}.

\begin{table}[htb!]
\renewcommand{\arraystretch}{0.90}
\parbox{0.45\linewidth}{
\centering
\begin{tabular}{|c|c|c|c|}
\hline
$\theta$ & $|a \vec{p}|^2$ & $Q^2 / M_\pi^2$ & $F_\pi(Q^2)$\\
\hline
\hline
0 & 0 & 0 & 1.000000(0)\\
\hline
0.0898 & 0.000414525 & 0.429212 & 0.9883(11)\\
\hline
0.1270 & 0.000829098 & 0.858473 & 0.9755(18)\\
\hline
0.16395 & 0.00138172 & 1.43068 & 0.9585(25)\\
\hline
0.2268 & 0.00264414 & 2.73782 & 0.9215(40)\\
\hline
0.2840 & 0.00414606 & 4.29295 & 0.8808(63)\\
\hline
0.32795 & 0.00552858 & 5.72446 & 0.8465(89)\\
\hline
0.36665 & 0.00691038 & 7.15521 & 0.811(12)\\
\hline
0.40165 & 0.00829266 & 8.58647 & 0.776(17)\\
\hline
0.4276 & 0.00939883 & 9.73183 & 0.749(22)\\
\hline
0.4864 & 0.0121615 & 12.5923 & 0.685(42)\\
\hline
\end{tabular}
\caption{\it \footnotesize Values of the angle $\theta$, the squared 3-momentum $|\vec{p}|^2$ in lattice units, the squared 4-momentum transfer $Q^2$ in units of the pion mass and the values of the pion form factor $F_\pi(Q^2)$ for the ensemble cA2.09.48. Errors are statistical only.}
\label{tab:table1}
}
\quad
\parbox{0.45\linewidth}{
\vspace{-1.75cm}
\centering
\begin{tabular}{|c|c|c|c|}
\hline
$\theta$ & $|a \vec{p}|^2$ & $Q^2 / M_\pi^2$ & $F_\pi(Q^2)$\\
\hline
\hline
0 & 0 & 0 & 1.000000(0)\\
\hline
0.11975 & 0.000414641 & 0.431681 & 0.98641(78)\\
\hline
0.2186 & 0.00138172 & 1.43851 & 0.9560(19)\\
\hline
0.3024 & 0.00264414 & 2.7528 & 0.9183(36)\\
\hline
0.37865 & 0.00414569 & 4.31606 & 0.8786(73)\\
\hline
0.43725 & 0.00552816 & 5.75534 & 0.846(13)\\
\hline
0.48885 & 0.00690991 & 7.19387 & 0.814(20)\\
\hline
0.5360 & 0.00830712 & 8.64851 & 0.782(27)\\
\hline
0.5701 & 0.00939773 & 9.78394 & 0.756(32)\\
\hline
0.64855 & 0.0121621 & 12.6619 & 0.685(54)\\
\hline
&&&\\
\hline
\end{tabular}
\caption{\it \footnotesize The same as in Table \ref{tab:table1} but for the ensemble cA2.09.64.}
\label{tab:table2}
}
\renewcommand{\arraystretch}{1.0}
\end{table}

\begin{table}[htb!]
\renewcommand{\arraystretch}{0.90}
\parbox{0.45\linewidth}{
\centering
\begin{tabular}{|c|c|c|c|}
\hline
$\theta$ & $|a \vec{p}|^2$ & $Q^2 / M_\pi^2$ & $F_\pi(Q^2)$\\
\hline
\hline
0 & 0 & 0 & 1.000000(0)\\
\hline
0.0449 & 0.000414525 & 0.125743 & 0.9882(20)\\
\hline
0.0635 & 0.000829098 & 0.251501 & 0.9781(25)\\
\hline
0.0820 & 0.00138257 & 0.419392 & 0.9655(31)\\
\hline
0.1134 & 0.00264414 & 0.802082 & 0.9387(48)\\
\hline
0.1420 & 0.00414606 & 1.25768 & 0.9089(70)\\
\hline
0.16395 & 0.0055269 & 1.67655 & 0.8831(91)\\
\hline
0.1833 & 0.00690849 & 2.09564 & 0.859(11)\\
\hline
0.2138 & 0.00939883 & 2.85107 & 0.819(14)\\
\hline
0.2432 & 0.0121615 & 3.68909 & 0.781(17)\\
\hline
0.27835 & 0.0159309 & 4.83254 & 0.737(23)\\
\hline
0.3154 & 0.0204542 & 6.20463 & 0.692(34)\\
\hline
0.3608 & 0.0267665 & 8.11943 & 0.645(53)\\
\hline
0.40205 & 0.0332368 & 10.0821 & 0.609(81)\\
\hline
0.4536 & 0.0423063 & 12.8333 & 0.57(13)\\
\hline
\end{tabular}
\caption{\it \footnotesize The same as in Table \ref{tab:table1} but for the ensemble cA2.30.24.}
\label{tab:table3}
}
\quad
\parbox{0.45\linewidth}{
\centering
\begin{tabular}{|c|c|c|c|}
\hline
$\theta$ & $|a \vec{p}|^2$ & $Q^2 / M_\pi^2$ & $F_\pi(Q^2)$\\
\hline
\hline
0 & 0 & 0 & 1.000000(0)\\
\hline
0.0898 & 0.000414525 & 0.132256 & 0.98901(69)\\
\hline
0.1270 & 0.000829098 & 0.264528 & 0.97806(93)\\
\hline
0.16395 & 0.00138172 & 0.440845 & 0.9638(12)\\
\hline
0.2268 & 0.00264414 & 0.843625 & 0.9327(17)\\
\hline
0.2840 & 0.00414606 & 1.32282 & 0.8981(27)\\
\hline
0.36665 & 0.00691038 & 2.20479 & 0.8399(53)\\
\hline
0.40165 & 0.00829266 & 2.64581 & 0.8128(70)\\
\hline
&&&\\
\hline
&&&\\
\hline
&&&\\
\hline
&&&\\
\hline
&&&\\
\hline
&&&\\
\hline
&&&\\
\hline
\end{tabular}
\caption{\it \footnotesize The same as in Table \ref{tab:table1} but for the ensemble cA2.30.48.}
\label{tab:table4}
}
\renewcommand{\arraystretch}{1.0}
\end{table}

\begin{table}[htb!]
\renewcommand{\arraystretch}{0.90}
\parbox{0.45\linewidth}{
\centering
\begin{tabular}{|c|c|c|c|}
\hline
$\theta$ & $|a \vec{p}|^2$ & $Q^2 / M_\pi^2$ & $F_\pi(Q^2)$\\
\hline
\hline
0 & 0 & 0 & 1.000000(0)\\
\hline
0.05985 & 0.000414295 & 0.066773 & 0.99116(76)\\
\hline
0.08465 & 0.000828772 & 0.133575 & 0.9818(11)\\
\hline
0.10935 & 0.00138299 & 0.2229 & 0.9694(14)\\
\hline
0.1512 & 0.00264414 & 0.426163 & 0.9420(20)\\
\hline
0.18935 & 0.00414679 & 0.668348 & 0.9113(27)\\
\hline
0.2186 & 0.0055269 & 0.890784 & 0.8847(33)\\
\hline
0.2444 & 0.00690849 & 1.11346 & 0.8598(40)\\
\hline
0.28505 & 0.00939773 & 1.51466 & 0.8184(53)\\
\hline
0.32425 & 0.0121602 & 1.95989 & 0.7773(68)\\
\hline
\end{tabular}
\caption{\it \footnotesize The same as in Table \ref{tab:table1} but for the ensemble cA2.60.32.}
\label{tab:table5}
}\quad
\parbox{0.45\linewidth}{
\centering
\begin{tabular}{|c|c|c|c|}
\hline
$\theta$ & $|a \vec{p}|^2$ & $Q^2 / M_\pi^2$ & $F_\pi(Q^2)$\\
\hline
\hline
0 & 0 & 0 & 1.000000(0)\\
\hline
0.0449 & 0.000414525 & 0.0658927 & 0.9886(12)\\
\hline
0.0635 & 0.000829098 & 0.131793 & 0.9785(16)\\
\hline
0.0820 & 0.00138257 & 0.219772 & 0.9655(20)\\
\hline
0.1134 & 0.00264414 & 0.420311 & 0.9379(27)\\
\hline
0.1420 & 0.00414606 & 0.659055 & 0.9074(37)\\
\hline
0.16395 & 0.0055269 & 0.878552 & 0.8810(47)\\
\hline
0.1833 & 0.00690849 & 1.09817 & 0.8558(58)\\
\hline
0.2138 & 0.00939883 & 1.49403 & 0.8130(80)\\
\hline
0.2432 & 0.0121615 & 1.93318 & 0.769(11)\\
\hline
\end{tabular}
\caption{\it \footnotesize The same as in Table \ref{tab:table1} but for the ensemble cA2a.60.24.}
\label{tab:table6}
}
\renewcommand{\arraystretch}{1.0}
\end{table}

%%%%%%%%%%%%%%%%%%%%%%%%%%%%%%%%%%%%%%%%%%%%%%%%
\bibliographystyle{h-physrev5}

\bibliography{bibliography}

\end{document}